\documentclass[final]{ias2}

\usepackage{graphicx}
\usepackage{multirow}
\usepackage{array}

\usepackage{hyperref}

\newcommand{\mev}{\,\mathrm{MeV}}

\newcommand{\mevm}{\mathrm{MeV}/c^2}

\newcommand{\gevm}{\mathrm{GeV}/c^2}

\newcommand{\pp}{\pi^+\pi^-}

\newcommand{\Uf}{\Upsilon(5S)}
\newcommand{\Uo}{\Upsilon(1S)}
\newcommand{\Un}{\Upsilon(nS)}
\newcommand{\Ut}{\Upsilon(2S)}
\newcommand{\Uth}{\Upsilon(3S)}

\newcommand{\mmpp}{MM(\pi^+\pi^-)}

\newcommand{\hb}{h_b(1P)}

\newcommand{\hbp}{h_b(2P)}
\newcommand{\hbn}{h_b(mP)}

\newcommand{\pipm}{\pi^{\pm}}

\newcommand{\zbo}{Z_b(10610)}
\newcommand{\zbt}{Z_b(10650)}

\newcommand{\mzahb}{10605.1\pm2.2\,^{+3.0}_{-1.0}}
\newcommand{\gzahb}{11.4\,^{+4.5}_{-3.9}\,^{+2.1}_{-1.2}}
\newcommand{\mzbhb}{10654.5\pm2.5\,^{+1.0}_{-1.9}}
\newcommand{\gzbhb}{20.9\,^{+5.4}_{-4.7}\,^{+2.1}_{-5.7}}
\newcommand{\ahb}{1.8\,^{+1.0}_{-0.7}\,^{+0.1}_{-0.5}}
\newcommand{\phihb}{188\,^{+44}_{-58}\,^{+4}_{-9}}

\newcommand{\mzahbp}{10596\pm7\,^{+5}_{-2}}
\newcommand{\gzahbp}{16\,^{+16}_{-10}\,^{+13}_{-4}}
\newcommand{\mzbhbp}{10651\pm4\pm2}
\newcommand{\gzbhbp}{12\,^{+11}_{-9}\,^{+8}_{-2}}
\newcommand{\ahbp}{1.3\,^{+3.1}_{-1.1}\,^{+0.4}_{-0.7}}
\newcommand{\phihbp}{255\,^{+56}_{-72}\,^{+12}_{-183}}

\begin{document}

\markboth{Pramana class file for \LaTeX 2e}{Hai-Bo Li}

\title{Light hadron, Charmonium(-like) and Bottomonium(-like) states}

\author[sin]{Hai-Bo Li}
\email{lihb@ihep.ac.cn}
\address[sin]{Institute of High Energy Physics, Beijing - 100049, China}

\begin{abstract}
Hadron physics represents the study of strongly interacting matter
in all its manifestations and the understanding of its properties
and interactions. The interest on this field has been revitalized by
the discovery of new light hadrons, charmonium- and bottomonium-like
states. In this talk I review the most recent experimental results
from different experiments.
\end{abstract}



\maketitle


\section{Introduction}

The modern theory of the strong force, which binds quarks inside
hadron matter, is quantum chromodynamics (QCD)~\cite{gross,poli}. In
the QCD framework only color-singlet states can exist due to
confinement, and only some combinations of color states produce an
attractive potential, leading to a bound state. Therefore, the field
of hadronic physics is the study of strong interaction in all
visible matters and understanding of fundamental questions in terms
of QCD. Recently the experimental situation in this field appeared
quite exciting. Especially, the situation changed dramatically in
the last ten years. After the Belle Collaboration claimed the
observation of a new and narrow resonance around 3.872 GeV/$c^2$,
decaying into $J/\psi \pi^+\pi^+$, named
$X(3872)$~\cite{first-belle}, many similar charmonium and
bottomonium states was found. Most of these states do not fit in the
standard charmonium/bottomonium model.

This review talk is devoted to the experimental signatures of these
hadron spectroscopy, in particular, these exotic states. These
results are from experiments at both $e^+ e^-$ and hadron colliders.
Most of the new charmonium states observed in the last few years are
from $B$ factory experiments BABAR and Belle. Many new observations
of hadron spectroscopy also come from the $\tau-$charm factory
experiments BESIII and CLEO-c, and from hadron colliders including
experiments at Tevatron and LHC, as well as fixed target
experiments, such as COMPASS. The analysis techniques adopted in the
search for hadron spectroscopy depend on the production mechanisms
at different experiments. In the $B$ factories at $e^+e^-$ machine,
the hadron can be produced in a $B$ meson decay, e.g. $B\rightarrow
K h$, and in the $s$-channel direct production, $e^+e^- \rightarrow
\gamma^* \rightarrow h$ or with an initial state radiation (ISR) for
hadron state carrying $J^{PC}=1^{--}$, and in the two photon fusion
production, as well as double quarkonium production, e.g. $e^+e^-
\rightarrow (c\bar{c})(c\bar{c})$. At hadron collider, prompt
productions coexist with the production in $B$ meson decays. At a
fixed-target experiment like COMPASS, the production mechanisms are
mainly from three mechanisms: central production, diffractive
dissociation and photoproduction. In this paper, for the reported
experimental results, the first error and second error will be
statistical and systematic, respectively, if they are not specified.

\section{Light meson decay and spectroscopy}

\subsection{$\eta$ and $\eta^\prime$ mesons and their excitation
states}

The $\eta$ and $\eta^\prime$ mesons play an important role in
understanding the low energy QCD. They are isoscalar members of the
nonet of the lightest pseudoscalar mesons.  Precision measurements
on $\eta$ and $\eta^\prime$ would be very helpful and provide useful
information in our understanding of low energy QCD.

\begin{figure}[ht]
\begin{center}
\includegraphics[width=0.35\columnwidth]{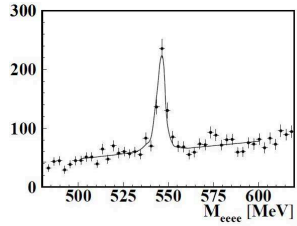}
\caption{$M_{e^+e^-e^+e^-}$ mass distribution from data and fitting
results.} \label{fig:eta-kloe}
\end{center}
\end{figure}

The conversion decay of $\eta \rightarrow e^+e^-e^+e^-$ is important
for the understanding of the $\eta$ coupling to the virtual photons
and the calculation of the anomalous magnetic moment of the
muon~\cite{land}. The KLOE Collaboration performed analysis based on
data sample of 1.7 fb$^{-1}$ collected at $\sqrt{s}=1.02$ GeV in the
$\phi$ meson mass region. By using the radiative decay $\phi
\rightarrow \gamma \eta$, the decay rate for the $\eta \rightarrow
e^+e^- e^+ e^- (\gamma)$  has been firstly obtained to be ${\mathcal
BR} (\eta \rightarrow e^+e^- e^+ e^- (\gamma)) = (2.4\pm 0.2 \pm
0.1) \times 10^{-5}$~\cite{kloe-eta}. As a result, the measured
branching ratio is fully radiation inclusive. The measurement is in
agreement with theoretical predictions, which are in the range
$(2.41-2.67)\times 10^{-5}$~\cite{jarlskog,miya,petri}.
Figure~\ref{fig:eta-kloe} shows the mass distribution of four
electrons.

Precise measurements of the $\eta$ and $\eta^\prime$ decay rates
provide important information to test the Chiral Perturbation Theory
(ChPT)~\cite{chpt}. The decays $\eta/\eta^\prime \rightarrow
\pi^+\pi^- \gamma$ are expected to get contributions from the box
anomalies which proceed through a vector meson resonance, described
by the Vector Meson Dominance (VDM)~\cite{vdm}. According to the
effective theory the $\eta/\eta^\prime \rightarrow \pi^+\pi^-
\gamma$ processes are supposed to proceed both via a resonant
contribution, mediated by the $\rho$ meson, and a non-resonant
direct term, connected to the box anomaly. Using a data sample
corresponding to an integrated luminosity of $558$ fb$^{-1}$ at
$\phi$ peak, the KLOE measured the ratio of $\eta \rightarrow
\pi^+\pi^-\gamma$ and $\eta \rightarrow \pi^+\pi^-\pi^0$ using the
decay $\phi \rightarrow \gamma \eta$~\cite{kloe-eta-2}. The
preliminary result is
\begin{equation}
\frac{\mathcal{BR}(\eta\rightarrow
\pi^+\pi^-\gamma)}{\mathcal{BR}(\eta\rightarrow \pi^+\pi^-\pi^0)} =
0.1838 \pm 0.0005\pm 0.0030. \label{eq:kloe-eta}
\end{equation}
The result is the most precision measurement and in agreement with
the recent results from CLEO~\cite{cleo-eta}, which differs by more
than 3$\sigma$ from the average of the previous
results~\cite{pdg2010}.

With new data sample of 225 million $J/\psi$ decay collected at the
BESIII detector, the $\eta$ and $\eta^\prime$ decays can be studied
via the charmonium decays into final states involving
$\eta/\eta^\prime$ meson. The rare and forbidden decays of $\eta$
and $\eta^\prime$ could be reached at the BESIII
experiment~\cite{li2009}. The BESIII Collaboration made a precision
measurement of $\eta^\prime \rightarrow 3 \pi$s via the decays
$J/\psi \rightarrow \gamma \pi^+\pi^-\pi^0$ and $\gamma
\pi^0\pi^0\pi^0$~\cite{besiii-eta1405}. The branching fractions are
determined to be $\mathcal{BR}(\eta^\prime \rightarrow
\pi^+\pi^-\pi^0)=(3.83\pm0.15 \pm0.39)\times 10^{-3}$ and
$\mathcal{BR}(\eta^\prime \rightarrow
\pi^0\pi^0\pi^0)=(3.56\pm0.22\pm0.34)\times 10^{-3}$. For the decay
$\eta^\prime \rightarrow \pi^+\pi^-\pi^0$, the branching ratio  is
consistent with the CLEO-c measurement~\cite{cleo-etap}, and the
precision is improved by a factor of four. For the decay
$\eta^\prime \rightarrow \pi^0\pi^0\pi^0$, it is two times larger
than the world average value~\cite{pdg2010}.
\begin{figure}[ht]
\begin{center}
\includegraphics[width=0.3\columnwidth]{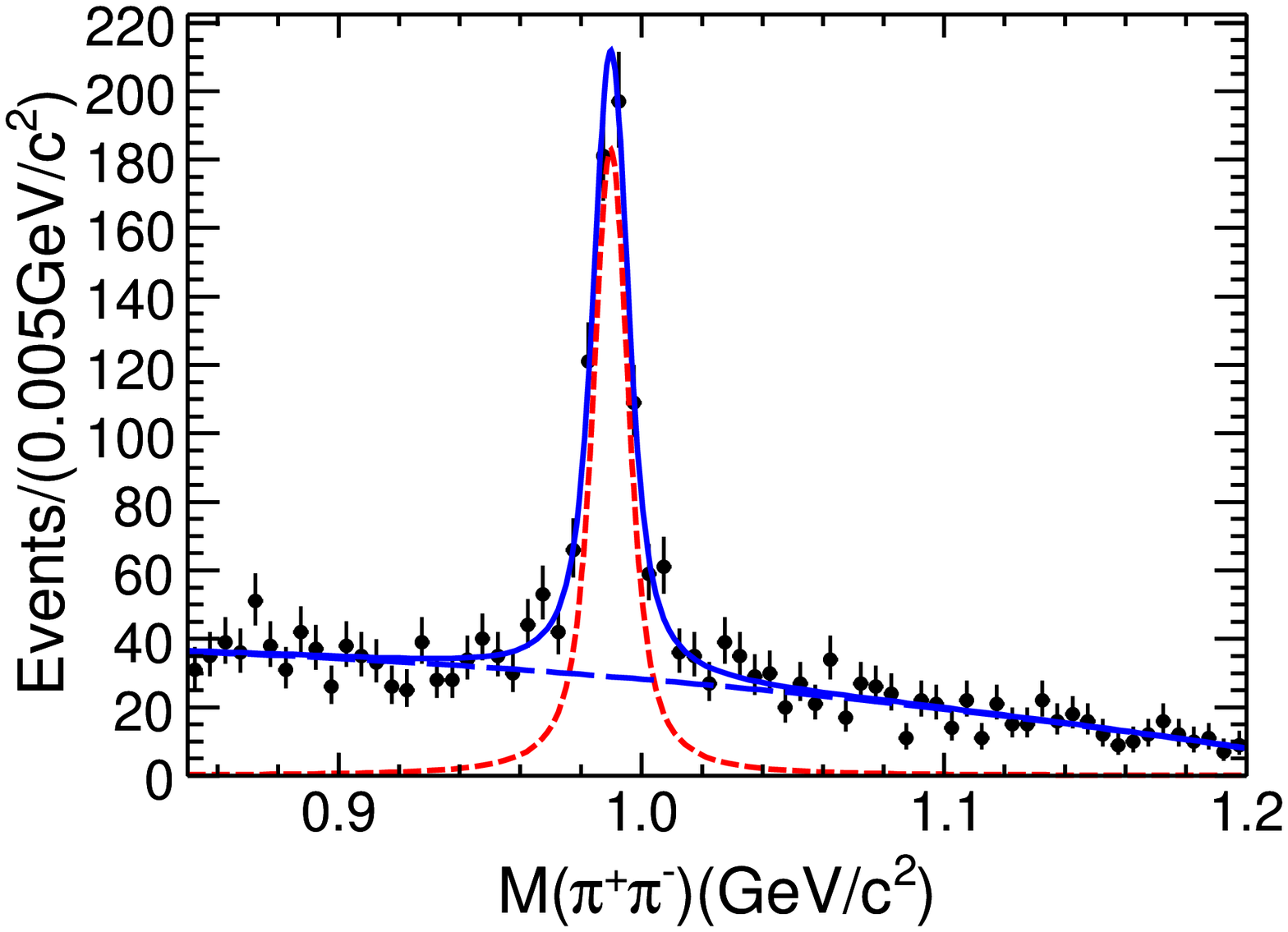}
\includegraphics[width=0.3\columnwidth]{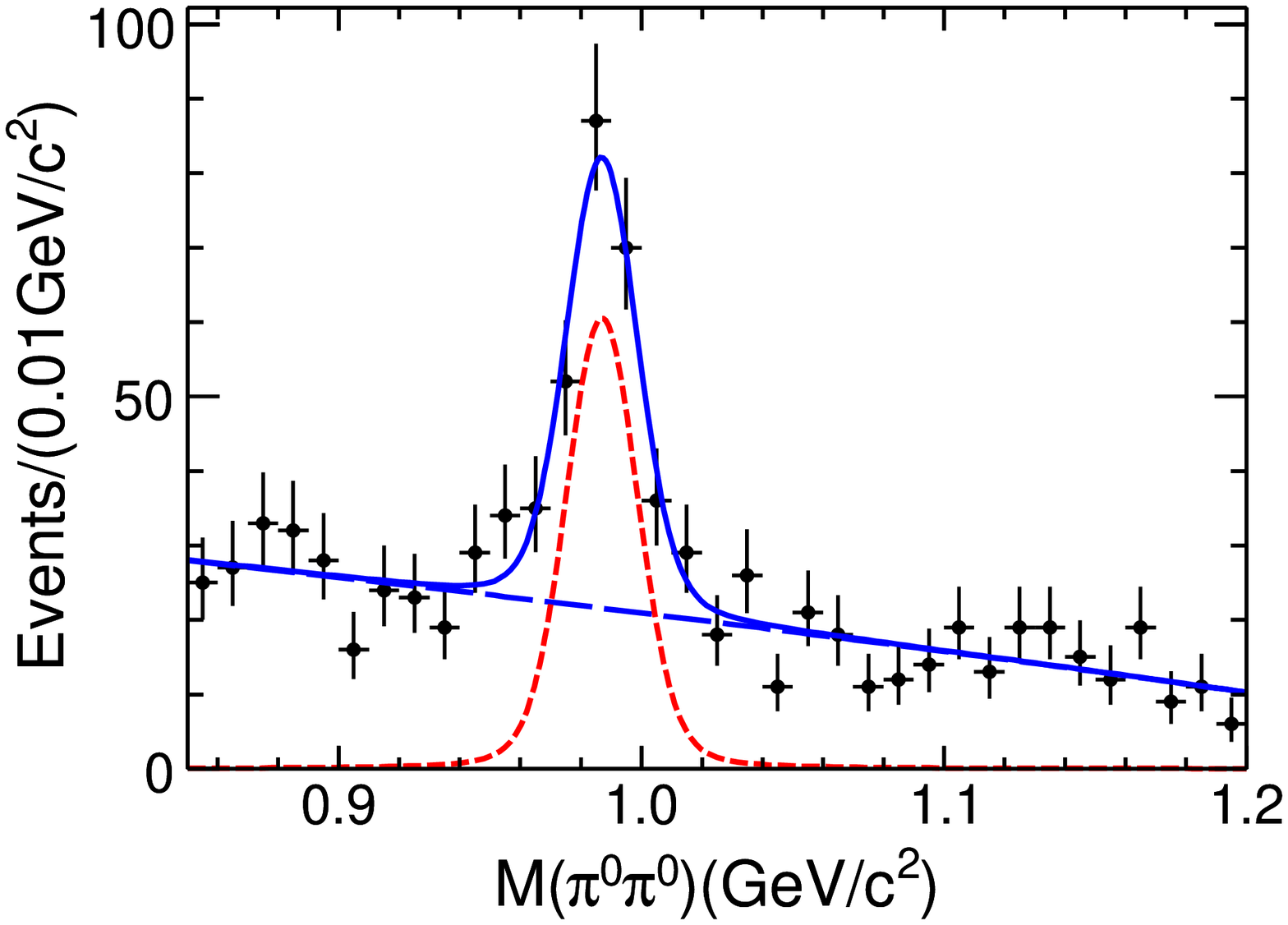}
\caption{The $\pi^+\pi^-$ (left) and $\pi^0\pi^0$ (right) invariant
mass spectra with $\pi^+\pi^-\pi^0$ ($3\pi^0$s) in the $\eta(1405)$
mass region. The solid curve is the result of the fit described in
the text. The dotted curve is the $f_0(980)$ signal. The dashed
curve denotes the background polynomial. } \label{fig:f0980}
\end{center}
\end{figure}
\begin{figure}[ht]
\begin{center}
\includegraphics[width=0.3\columnwidth]{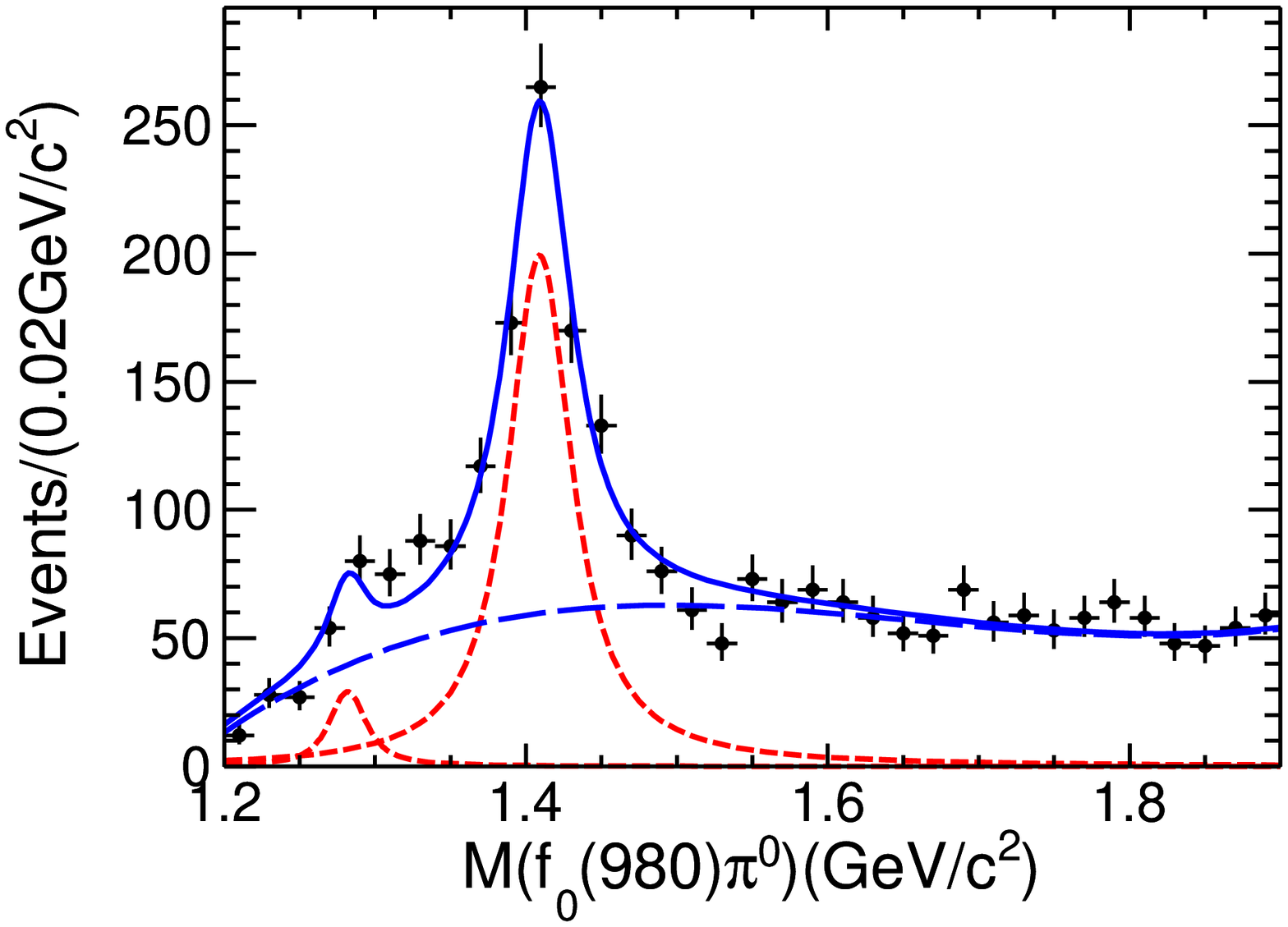}
\includegraphics[width=0.3\columnwidth]{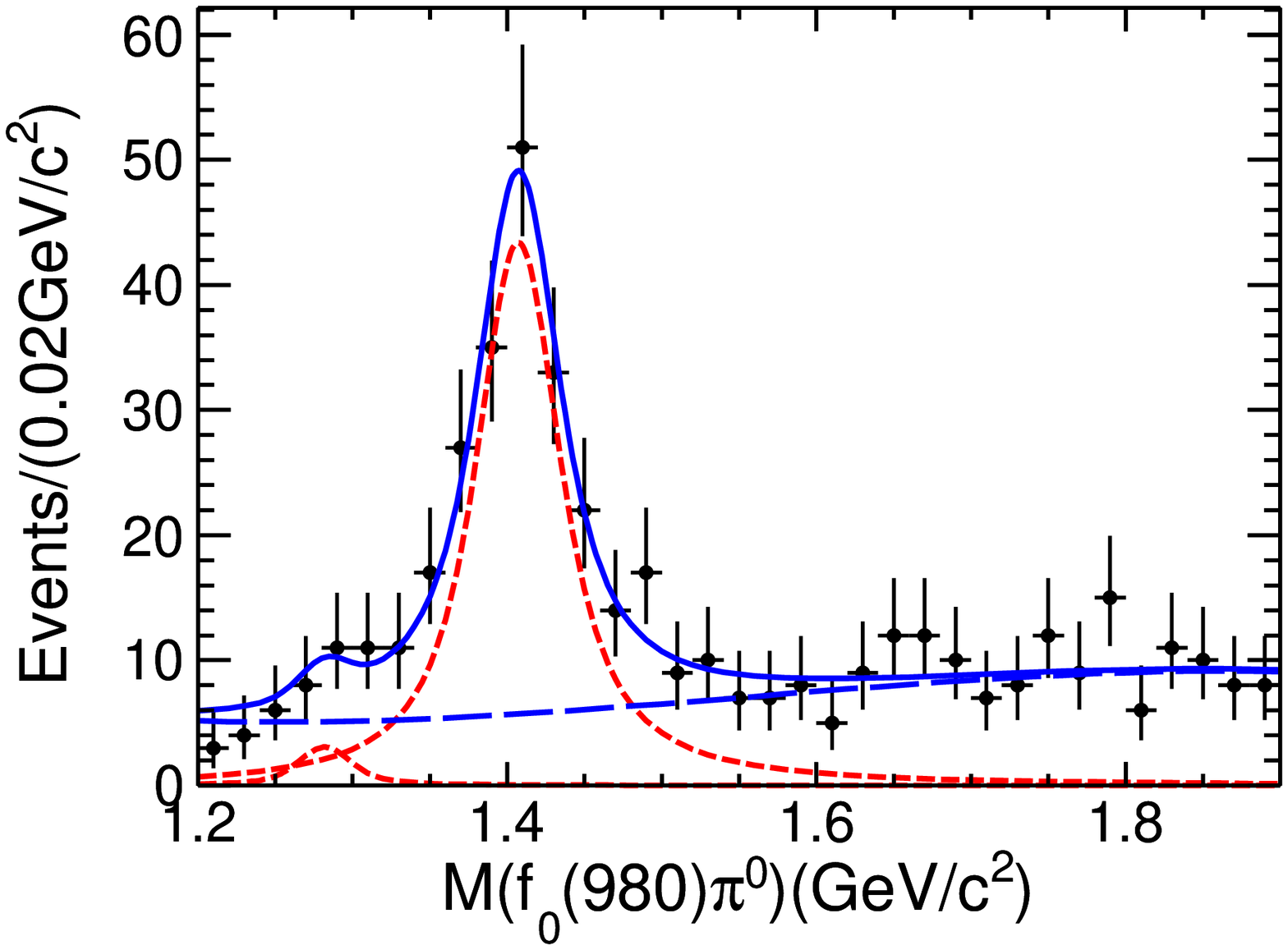}
\caption{Results of the fit to (left) the
$f_0(980)(\pi^+\pi^-)\pi^0$ and (right) $f_0(980)(\pi^0\pi^0)\pi^0$
invariant mass spectra. The solid curve is the result of the fit
described in the text. The dotted curve is the
$f_1(1285)/\eta(1295)$ and $\eta(1405)$ signal. The dashed curves
denote the background polynomial. } \label{fig:eta1405}
\end{center}
\end{figure}

The spectrum of radial excitation states of isoscalar $\eta$ and
$\eta^\prime$ is still not well known. An important issue is about
the nature of $\eta(1405)$ and $\eta(1475)$ states, which are not
well established. By using the decays of $J/\psi \rightarrow \gamma
\pi^+\pi^-\pi^0$ and $\gamma \pi^0\pi^0\pi^0$ in BESIII, clear
$f_0(980)$ signals are observed on both $\pi^+\pi^-$ and
$\pi^0\pi^0$ spectra as shown in Fig.~\ref{fig:f0980}, the width of
the observed $f_0(980)$ is much narrower (less than 10 MeV) than
that in other processes~\cite{pdg2010}. By taking events in the
window of $f_0(980)$ on the $\pi^+\pi^-$ ($\pi^0\pi^0$) mass
spectrum, evidence of $f_1(1285)/\eta(1295)$ is observed in the low
mass region of $f_0(980)\pi^0$ as shown in
Fig.~\ref{fig:eta1405}~\cite{besiii-eta1405}. It is interesting that
a clear peak around 1400 MeV is also observed on the mass of
$f_0(980)\pi^0$ (see Fig.~\ref{fig:eta1405}). Angular analysis
indicates that the peak on 1400 MeV is from $\eta(1405)\rightarrow
f_0(980)\pi^0$ decay. The BESIII Collaboration measured the product
branching fraction of $\eta(1405)$ production to be
$\mathcal{BR}(J/\psi \rightarrow \gamma \eta(1405))\times
\mathcal{BR}(\eta(1405) \rightarrow f_0(980)\pi^0) \times
\mathcal{BR}( f_0(980)\rightarrow  \pi^+\pi^-) = (1.50 \pm 0.11 \pm
0.11)\times 10^{-5}$ and $\mathcal{BR}(J/\psi \rightarrow \gamma
\eta(1405))\times \mathcal{BR}(\eta(1405) \rightarrow f_0(980)\pi^0)
\times \mathcal{BR}( f_0(980)\rightarrow \pi^0\pi^0) = (7.10\pm0.82
\pm0.72)\times 10^{-6}$~\cite{besiii-eta1405}, respectively.
According to PDG values~\cite{pdg2010}, one obtains the ratio
$\mathcal{BR}(\eta(1405)\rightarrow f_0(980)
\pi^0)/\mathcal{BR}(\eta(1405)\rightarrow a_0(980) \pi^0)\sim 25\%$.
 It is the first time that we observe anomalously
large isospin violation in the strong decay of $\eta(1405)
\rightarrow f_0(980)\pi^0$ (for example, the ratio of the isospin
violating $\eta^\prime \rightarrow \pi^+\pi^-\pi^0$ to the isospin
covering $\eta^\prime \rightarrow \pi^+\pi^- \eta$ is about 0.8\%).
Following BESIII measurement, in reference~\cite{qiang}, the authors
interpret this puzzle as an intermediate on-shell $K\bar{K}^* +
c.c.$ rescattering to the isospin violating $f_0(980)\pi^0$ by
exchanging on-shell kaon. Further experimental study on the
$\eta(1405)$ parameters and identification of quantum number are
needed at BESIII.

\subsection{Spin exotic light states}

COMPASS is a multi-purpose fixed-target experiment at the CERN Super
Proton Synchrotron (SPS) aimed at studying the structure and
spectrum of hadrons. One primary goal is the search for the
spin-exotic mesons and glueballs. In a partial-wave analysis (PWA)
of the pilot run data taken in 2004, a significant spin-exotic
$J^{PC} = 1^{-+}$ resonance was found at around 1660 MeV/$c^2$ in
$\pi^-\pi+\pi^-$ final states produced in $\pi^-$ diffraction on a
$Pb$ target~\cite{compass2010}. Its mass-dependent phase differences
to the $J^{PC} = 2^{-+}$ and $1^{++}$ waves are consistent with the
highly debated $\pi_1(1600)$ meson claimed in this channel by E852
and VES experiments~\cite{e852,ves}. From a mass-dependent fit a
resonance mass of $(1660\pm 10^{+0}_{-64})$ MeV/$c^2$ and a width of
$(269\pm 21^{+42}_{-64})$ MeV/$c^2$ are deduced.
\begin{figure}[ht]
\begin{center}
\includegraphics[width=0.8\columnwidth]{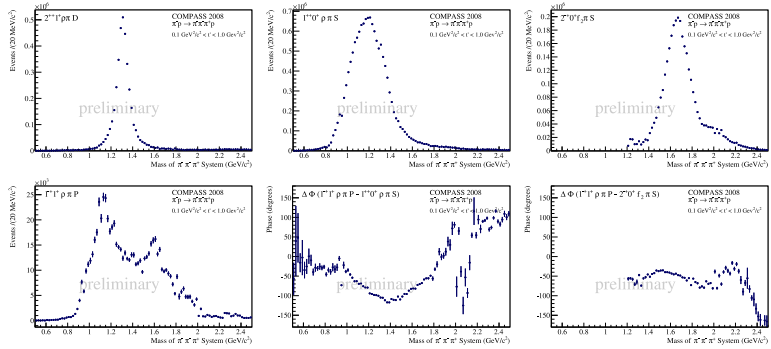}
\caption{{\it Top row}: Intensities of major waves:
$2^{++}1^+[\rho\pi]D$ with $a_2(1320)$ (left), $1^{++}0^+[\rho\pi]S$
with $a_1(1260)$ (center) and $2^{-+}0^+ [f_2\pi]S$ with
$\pi_2(1670)$ (right). {\it Bottom row}: Intensity of the
spin-exotic wave $1^{-+}1^+[\rho\pi]P$ (left), and phase differences
of this wave with respect to the $1^{++}0^+[\rho\pi]S$ (center), and
the $2^{-+}0^+ [f_2\pi]S$ waves (right) (From~\cite{boris}).}
\label{fig:compass-1}
\end{center}
\end{figure}

In 2008 COMPASS has acquired large data sets of diffractive
dissociation of 190 GeV/c $\pi^-$ on a H$_2$ target. In a
partial-wave analysis (PWA) the isobar model~\cite{isobar} is used
to decompose the decay $X^- \rightarrow \pi^-\pi^+\pi^-$ into a
chain of successive two-body decays. The spin-density matrix is
determined by extended maximum likelihood fits performed in 20
MeV/$c^2$ wide bins of the three-pion invariant mass. The intensity
of the three dominant waves in the $\pi^-\pi^+\pi^-$ final state,
$1^{++} 0^+ [\rho\pi]S$ , $2^{++} 1^+ [\rho\pi]D$, and $2^{-+} 0^+ [
f_2\pi ]S$ , are shown in Fig.~\ref{fig:compass-1}, top row. They
contain resonant structures that correspond to the $a_1(1260)$,
$a_2(1320)$, and $\pi_2(1670)$, respectively~\cite{boris}. Of
peculiar interest are the fit results for the spin-exotic wave in
Fig.~\ref{fig:compass-1} bottom, left plot. The plot nicely
illustrates the unprecedented statistical accuracy due to the large
data set. The $1^{-+}1^+ [\rho\pi]P$ intensity  features a broad
bump, centered at 1.6 GeV/$c^2$. In this mass region a rising phase
with respect to the tail of the $a_1(1260)$ in the $1^{++} 0^+
[\rho\pi]S$ wave is seen (Fig.~\ref{fig:compass-1} bottom, center).
 As in Fig. 1 (bottom, right shows), the structure is phase locked with the
$\pi_2(1670)$ in the $2^{-+}0^+ [ f_2\pi ]S$ wave. This is
consistent with the results obtained from a PWA of the pilot-run
data taken with a $Pb$ target~\cite{compass2010}.

The CLEO-c Collaboration presents an analysis  of $\psi(2S)
\rightarrow \gamma \chi_{c1} \rightarrow \gamma \eta^{(\prime)}
\pi^+\pi^-$ decays in which they study the production of various
$\eta^{(\prime)}\pi$ intermediate states~\cite{cleo-c-pi1}. They
find evidence for an exotic $\eta^\prime \pi$ $P$-wave scattering
amplitude at the level of 4 standard deviations under a wide variety
of model variations. The best fit to the data is obtained when the
$\pi_1 \pi$ amplitude is included. The $\eta^\prime \pi$ mass is
described by a Breit-Wigner lineshape with a mass and width of $1670
\pm 30 \pm 20$ MeV/$c^2$ and $240\pm50\pm60$ MeV/$c^2$,
respectively, which is the first evidence of exotic state in
charmonium decays.  The result is consistent with that from other
experiments~\cite{e852,ves}. Now BESIII had collected 4 times of
CLEO-c's sample at $\psi(2S)$ peak~\cite{lihbhadron2011}, we expect
to confirm CLEO-c measurement soon.

\subsection{New light mesons observed by BESIII}

\begin{figure}[ht]
\begin{center}
\includegraphics[width=0.3\columnwidth]{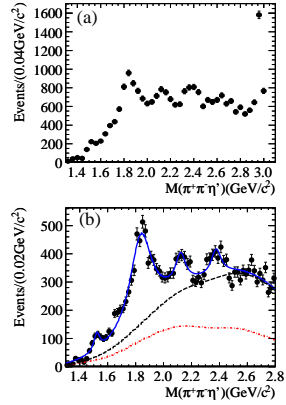}
\caption{(a) The $\eta^\prime \pi^+\pi^-$ invariant-mass
distribution. (b) mass spectrum fitting with four resonances, here,
the dash-dot line is contributions of non-$\eta^\prime$ events and
the $\eta^\prime \pi^+\pi^-\pi^0$ background and the dash line is
contributions of the total background and non-resonant $\eta^\prime
\pi^+\pi^-$ process.}
 \label{fig:x1835}
\end{center}
\end{figure}

In 2005, BESII observed a $\eta^\prime \pi^+\pi^-$ resonance,
$X(1835)$, in the radiative decay $J/\psi \rightarrow \gamma
\eta^\prime \pi^+\pi^-$ with a statistical significance of
7.7$\sigma$~\cite{besii1835}. A fit to a Breit-Wigner function
yielded a mass $M = 1833.7 \pm 6.1 \pm 2.7$ MeV/$c^2$, a width
$\Gamma = 67.7 \pm 20.3 \pm 7.7$ MeV/$c^2$. The study was stimulated
by searching for $0^{-+}$ glueball candidates which was predicted by
the Lattice QCD~\cite{lqcd}. With 225 million $J/\psi$ decay events
collected by the BESIII detector, the $X(1835)$ state has been
confirmed with a statistical significance larger than 20$\sigma$ in
the same analysis~\cite{besiii1835}. The mass and width are measured
to be $M = 1836.5 \pm 3.0^{+5.6}_{-2.1}$ MeV/$c^2$ and $\Gamma = 190
\pm 9^{+38}_{-36}$ MeV/$c^2$ as shown in shown in
Fig.~\ref{fig:x1835}. The mass of the $X(1835)$ is consistent with
the BESII result, but the width is significantly larger. A simple
angular analysis indicates that quantum number of $X(1835)$ is
consistent with a pseudoscalar assignment, but the others are not
excluded.
Meanwhile, two resonances, the $X(2120)$ and the $X(2370)$ are
observed with statistical significances larger than 7.2$\sigma$ and
$6.4\sigma$, respectively, as shown in Fig.~\ref{fig:x1835}. The
masses and widths are measured to be:
\begin{itemize}
\item $X(2120)$
\begin{equation}
M = 2122.4\pm 6.7^{+4.7}_{-2.7}\, \text{MeV}/c^2, \,\, \Gamma=83\pm
16^{+31}_{-11}\, \text{MeV}/c^2.
 \label{eq:x2120}
\end{equation}
\item $X(2370)$
\begin{equation}
M = 2376.3\pm 8.7^{+3.2}_{-4.3}\, \text{MeV}/c^2, \,\, \Gamma=83\pm
17^{+44}_{-6}\, \text{MeV}/c^2.
 \label{eq:x2370}
\end{equation}
\end{itemize}
In the mass spectrum fitting in Fig.~\ref{fig:x1835}(b), possible
interferences among different resonances and the non- resonant
process are not taken into account. To determine the spin and parity
of the $X(1835)$, $X(2120)$ and $X(2370)$, and to measure their
masses and widths more precisely, a PWA must be performed, which
will be possible with the much higher statistics $J/\psi$ data
samples planned for future runs of the BESIII experiment.

\begin{figure}[ht]
\begin{center}
\includegraphics[width=0.4\columnwidth]{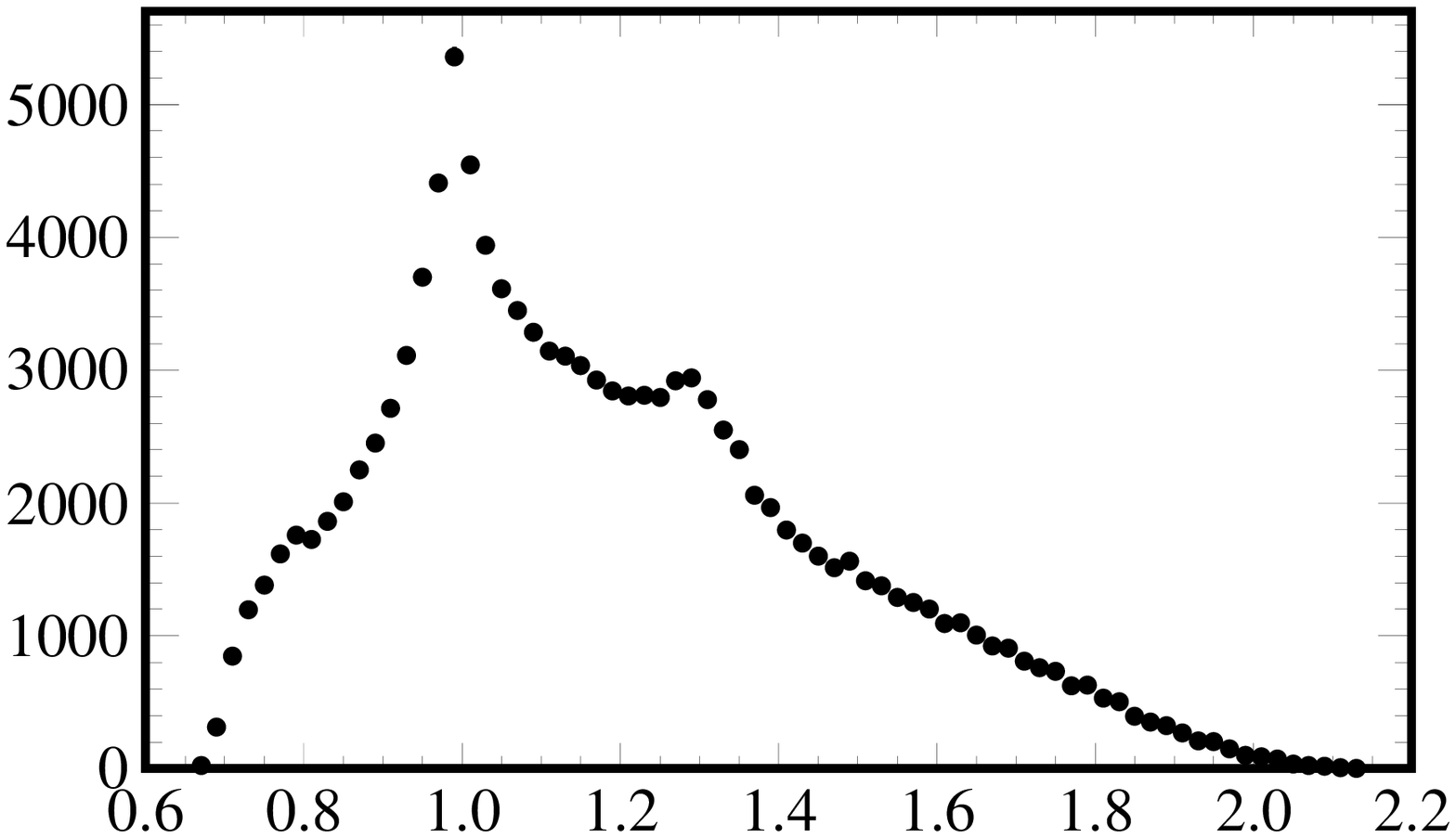}
\includegraphics[width=0.4\columnwidth]{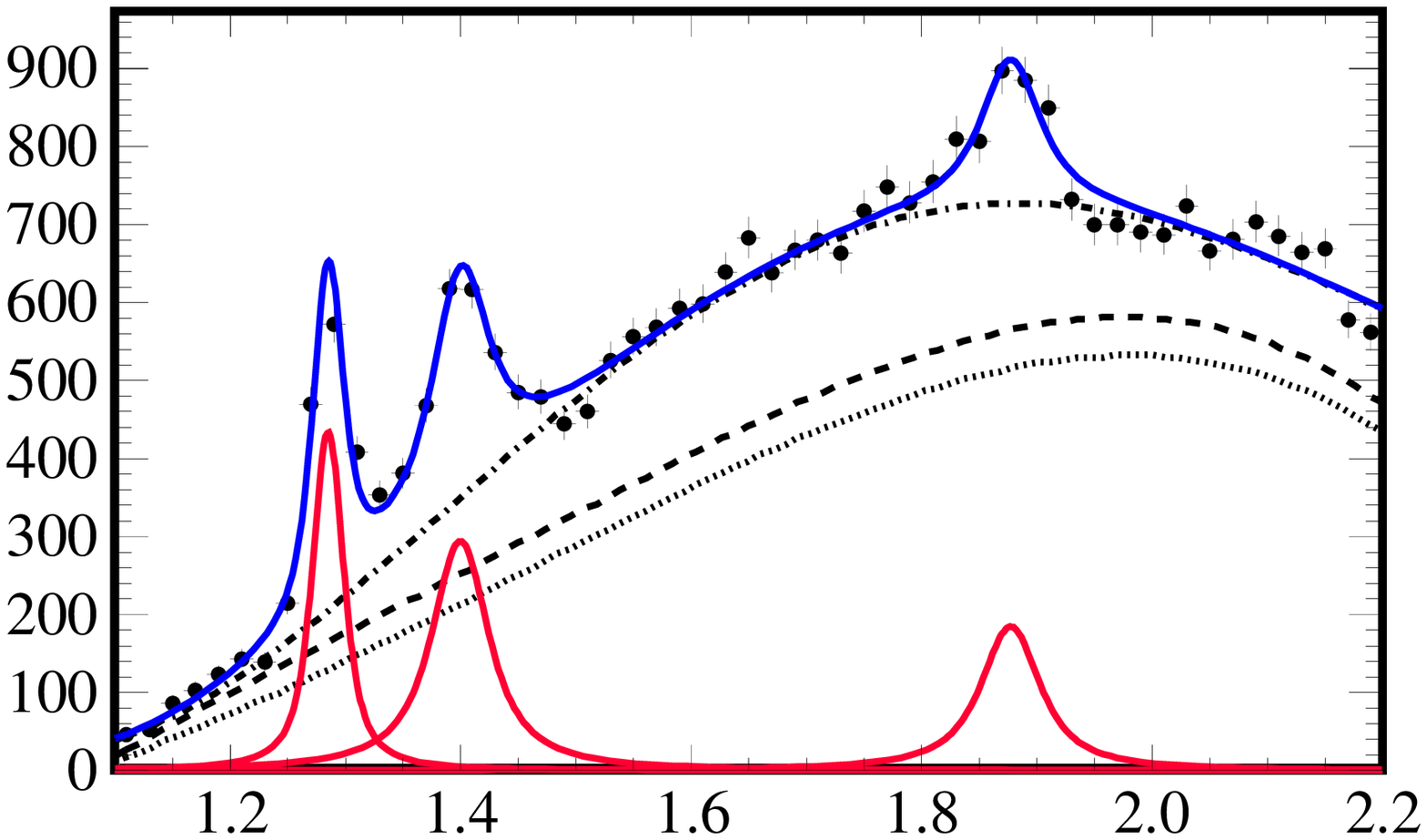}
\caption{ Invariant-mass distributions for the selected events: left
is the invariant-mass spectrum of $\eta \pi$; right: Results of the
fit to the $\eta \pi^+\pi^-$  mass distribution for events with
either $\eta\pi^+$ or $\eta\pi^-$ is in the $a_0(980)$ mass window.
 }
\label{fig:1870}
\end{center}
\end{figure}

The BESIII experiment also reports an analysis of $J/\psi
\rightarrow \omega \eta \pi^+\pi^-$. A structure around 1.8-1.9
GeV/$c^2$ in the $\eta \pi^+\pi^-$ mass spectrum is
observed~\cite{besiii1870}. In the analysis, a clear $a_0(980)$
signal is seen in the $\eta \pi$ mass spectrum as shown in
Fig.~\ref{fig:1870} (left). The  $\eta \pi^+\pi^-$ mass spectrum for
events where either $M(\eta\pi^+)$ or $M(\eta\pi^-)$ is in a $100$
MeV/$c^2$ mass window centered on the $a_0(980)$ mass is shown in
Fig.~\ref{fig:1870} (right).  Three peaks are observed on the $\eta
\pi^+\pi^-$ mass spectrum, two of them in the low mass side are
$f_1(1285)$ and $\eta(1405)$. The structure near 1.8 GeV/$c^2$ is
the first observation. A fit to the three signal peaks leads to $M=
1877.3\pm6.3^{+3.4}_{-7.4}$ MeV/$c^2$ and $\Gamma =
57\pm12^{+19}_{-4}$ MeV/$c^2$ for the $X(1870)$ structure with the
statistical significance of  7.2$\sigma$. Here the three signal
peaks are parametrized by Breit-Wigner functions convolved with a
Gaussian resolution function and multiplied by an efficiency curve,
which are both determined from signal MC samples and fixed in the
fit. Whether the resonant structure of $X(1870)$ is due to the
$X(1835)$, the $\eta_2(1870)$, an interference of both, or a new
resonance still needs further study such as a PWA that will be
possible with the larger $J/\psi$ data sample.

\subsection{Proton-anti-proton mass threshold enhancement}

\begin{figure}[ht]
\begin{center}
\includegraphics[width=0.4\columnwidth]{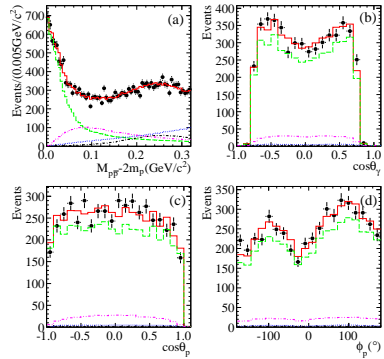}
\caption{Comparisons between data and PWA fit projection: (a) the
$p\bar{p}$ invariant mass; (b)-(d) the polar angle $\theta_\gamma$
of the radiative photon in the $J/\psi$  center of mass system, the
polar angle $\theta_p$ and the azimuthal angle $\phi_p$ of the
proton in the $p\bar{p}$ center of mass system with $M_{p\bar{p}} -
2M_p< 50$ MeV/$c^2$, respectively. Here, the black dots with error
bars are data, the solid histograms show the PWA total projection,
and the dashed , dotted , dash-dotted and dash-dot-dotted lines show
the contributions of the $X(p\bar{p})$, $0^{++}$ phase space,
$f_0(2100)$ and $f_2(1910)$, respectively. } \label{fig:ppbar-1}
\end{center}
\end{figure}

An strong $p\bar{p}$ mass threshold enhancement was first observed
by the BESII experiment in the decay process $J/\psi \rightarrow
\gamma p\bar{p}$~\cite{besiippbar} and was confirmed by the CLEO-c
experiment~\cite{cleocppbar}.  With the BESIII data, a PWA analysis
is performed to determine the parameters of the $p\bar{p}$ mass
threshold structure, which we denote as
$X(p\bar{p})$~\cite{besiiippbar}. Only events of $M_{p\bar{p}}< 2.2
$ GeV/$c^2$ are considered in the PWA.
 Four
components, the $X(p\bar{p})$, $f_2(1910)$, $f_0(2100)$ and $0^{++}$
phase space (PS) are included in the PWA fit. The intermediate
resonances are described by Breit-Wigner  propagators, and the
parameters of the $f_2(1910)$ and $f_0(2100)$ are fixed at PDG
values. In the fit, the $p\bar{p}$ final state interaction (FSI)
effect is also considered by using the Julich
formulation~\cite{fsi}. Figure~\ref{fig:ppbar-1} shows comparisons
of the mass and angular distributions between the data and the PWA
fit projections. In the optimal PWA fit, the $X(p\bar{p})$ is
assigned to be a $0^{-+}$ state, which is 6.8$\sigma$ better than
other $J^{PC}$ assignment. In the fit, a Breit-Wigner and S-wave
final state interaction (I=0) factor can well describe the
$p\bar{p}$ mass threshold structure. The PWA fits are also performed
without the correction for FSI effect. The corresponding
log-likelihood value worsen by 25.6 than those with FSI effect
included.
With the inclusion of Julich-FSI effects, the mass, width and
product BR for the $X(p\bar{p})$ are measured to be:
$M=1832^{+19}_{-5}(\text{stat.})^{+18}_{-17}(\text{syst.}) \pm
19(\text{model})$ MeV/$c^2$, $\Gamma = 13\pm 39
(\text{stat.})^{+10}_{-13}(\text{syst.}) \pm 4(\text{model})$
 MeV/$c^2$ ( a total width of $\Gamma < 76 $ MeV/$c^2$ at 90\% C.L.)
 and $\mathcal{BR}(J/\psi \rightarrow \gamma
 X)\mathcal{BR}(X\rightarrow p\bar{p}) = (9.0^{+0.4}_{-1.1}(\text{stat.})^{+1.5}_{-5.0}(\text{syst.}) \pm
2.3(\text{model})) \times 10^{-5}$, respectively, where the third
error are uncertainty due to choosing different model of FSI
effect~\cite{fsi,fsi-1,fsi-2}.

\begin{figure}[ht]
\begin{center}
\includegraphics[width=0.4\columnwidth]{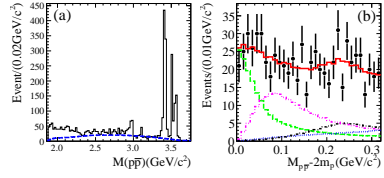}
\caption{(a) The $p\bar{p}$ invariant mass spectrum for the selected
$\psi(2S)\rightarrow \gamma p\bar{p}$ candidate events; the open
histogram is data and the dashed line is from a $\psi(2S)\rightarrow
\gamma p\bar{p}$ phase-space MC events(with arbitrary
normalization). (b) Comparisons between data and PWA fit projection
for $p\bar{p}$ mass spectrum, the representations of the error bars
and histograms are same as those in Fig.~\ref{fig:ppbar-1}.}
\label{fig:ppbar-2}
\end{center}
\end{figure}

The $\psi(2S)\rightarrow \gamma p\bar{p}$ decay is also studied by
using 106 million $\psi(2S)$ decay events collected by the BESIII
detector~\cite{besiiippbar}.  The $p\bar{p}$ mass spectrum of the
surviving events is shown in Fig.~\ref{fig:ppbar-2}(a). Besides the
well known $\eta_c$ and $\chi_{cJ}$ peaks, there is also a
$p\bar{p}$ mass threshold excess relative to phase space. However,
here the line shape of the mass spectrum in the threshold region
appears to be less pronounced than that in $J/\psi$ decays. A PWA on
the selected $\psi(2S) \rightarrow \gamma p\bar{p}$ which is similar
to that applied for $J/\psi \rightarrow \gamma p\bar{p}$ decay was
performed to check the contribution of $X(p\bar{p})$ in $\psi(2S)$
decays and to measure the production ratio between $J/\psi$ and
$\psi(2S)$ radiative decays, $R = \mathcal{BR}(\psi(2S) \rightarrow
\gamma X(p\bar{p}))/\mathcal{BR}(J/\psi \rightarrow \gamma
X(p\bar{p}))$. Due to limited statistics of $\psi(2S)$ events, in
the PWA, the mass and width of $X(p\bar{p})$ as well as its $J^{PC}$
were fixed to the results obtained from $J/\psi$ decays.
Figure~\ref{fig:ppbar-2} (b) shows comparisons between data and MC
projections for the $p\bar{p}$ mass spectrum. The determined product
$\mathcal{BR}$ and $R$ value are $\mathcal{BR}(\psi(2S) \rightarrow
\gamma
 X)\mathcal{BR}(X\rightarrow p\bar{p}) = (4.57\pm0.36(\text{stat.}
 )^{+1.23}_{-4.07}(\text{syst.}) \pm
1.28(\text{model})) \times 10^{-6}$ and $R=
(5.08^{+0.71}_{-0.45}(\text{stat.})^{+0.67}_{-3.58}(\text{syst.})
\pm 0.12(\text{model}))\%$, respectively. It is suppressed compared
with the so called ¡°12\% rule¡±.

\section{Charmonium spectroscopy and decays}

\subsection{$\eta_c(1S)$ and $\eta_c(2S)$}
\subsubsection{$\eta_c(1S)$ resonance via $\psi(2S) \rightarrow
\gamma \eta_c$ decay}

Precise measurement of M1 transition of $\psi(2S)$ is important for
us to understand the QCD in the relativistic and nonperturbative
regimes.  The $\psi(2S) \rightarrow \gamma \eta_c$ transition is
also a source of information on the $\eta_c$ mass and width. There
is currently a 3.3$\sigma$ inconsistency in previous $\eta_c$ mass
measurements from $J/\psi$ and $\psi(2S)\rightarrow \gamma \eta_c$
(averaging $2977.3 \pm 1.3$ MeV/$c^2$) compared to $\gamma \gamma$
or $p\bar{p}$ production (averaging 2982.6 $\pm$1.0
MeV/$c^2$)\cite{pdg2010}. The width measurements also spread from 15
to 30 MeV, it is around 10 MeV in the earlier days of experiments
using $J/\psi$ radiative transition~\cite{mark3,bes1}, while the
recent experiments, including photon-photon fusion and $B$ meson
decays, gave higher mass and much wider
width~\cite{cleo-c,babar,belle1,belle2}. The recent study by the
CLEO-c experiment~\cite{cleo-c-new}, using both $\psi(2S)$ and
$J/\psi \rightarrow \gamma \eta_c$ decays, and pointed out there was
a distortion of the $\eta_c$ line shape. The CLEO-c attributed the
$\eta_c$ line-shape distortion to the energy-dependence of the M1
transition matrix element. In the $J/\psi \rightarrow \gamma \eta_c$
from CLEO-c, the distorted $\eta_c$ lineshape can be described by
the relativistic Breit-Wigner (BW) distribution modified by a factor
of E$^3_{\gamma}$ together with a dumping factor to suppress the
tail on the higher photon energies. The KEDR Collaboration did the
same thing but tried different dumping factor~\cite{kedr}.

\begin{figure}[ht]
\begin{center}
\includegraphics[width=0.3\columnwidth]{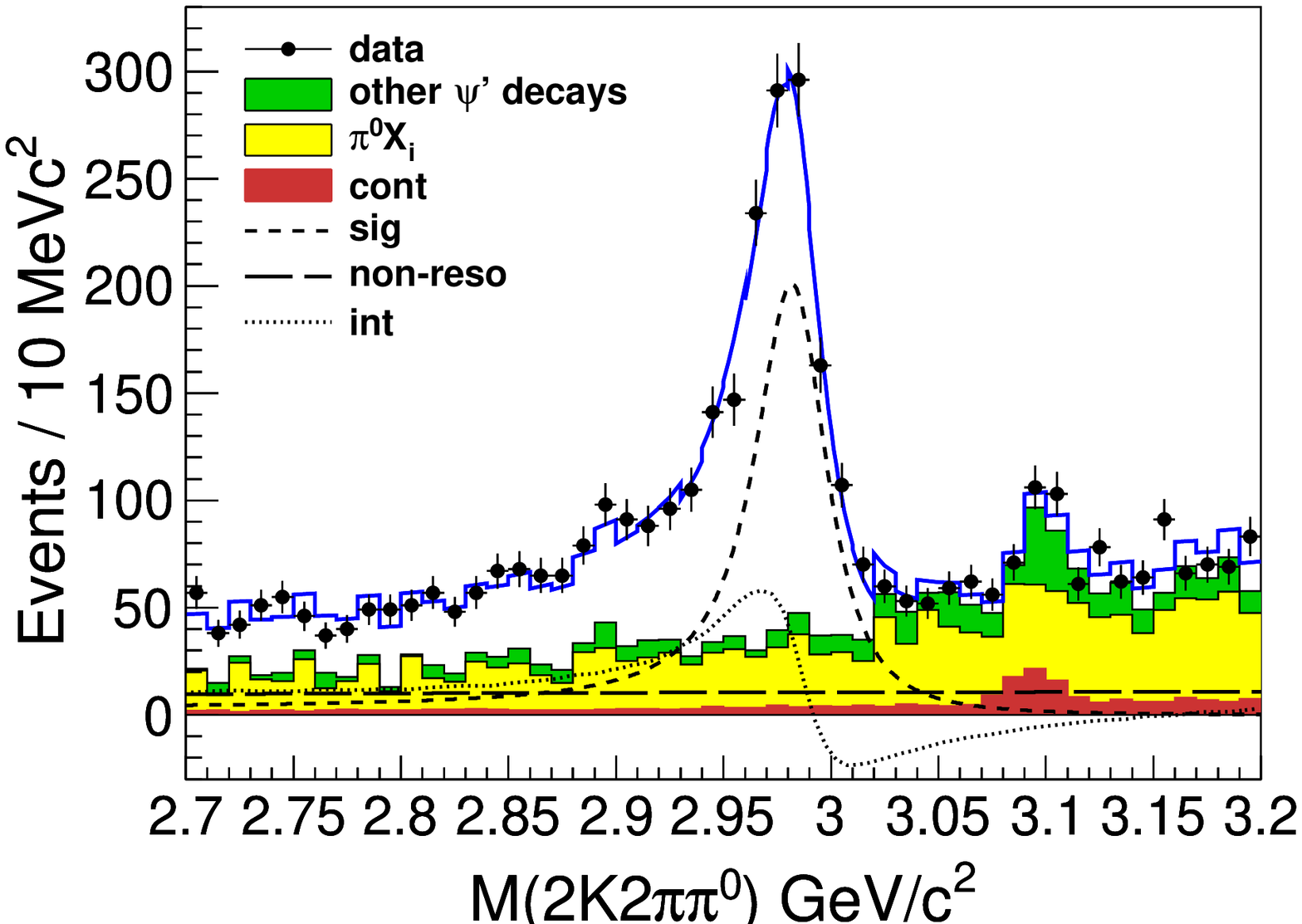}
\includegraphics[width=0.3\columnwidth]{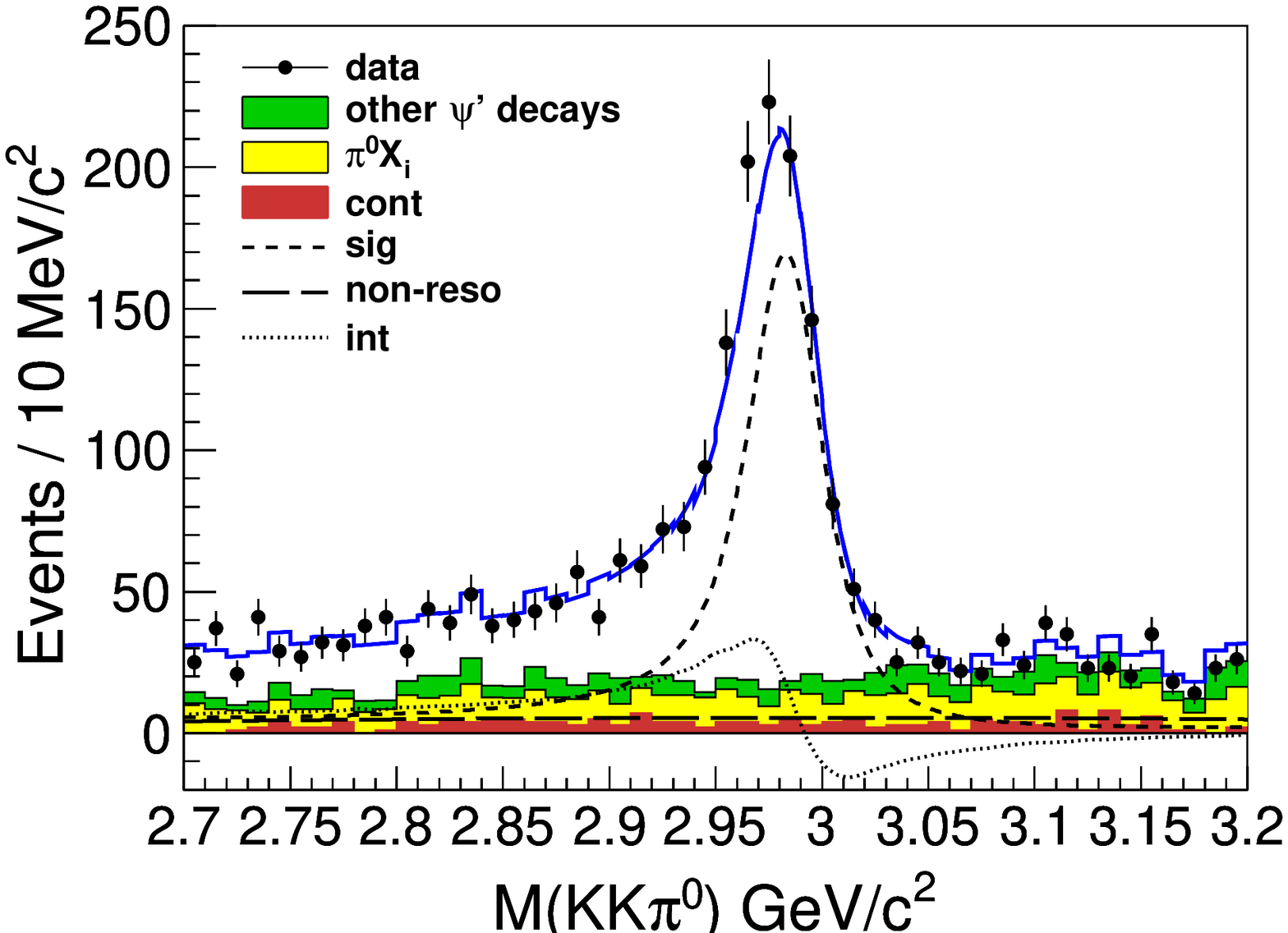}
\includegraphics[width=0.3\columnwidth]{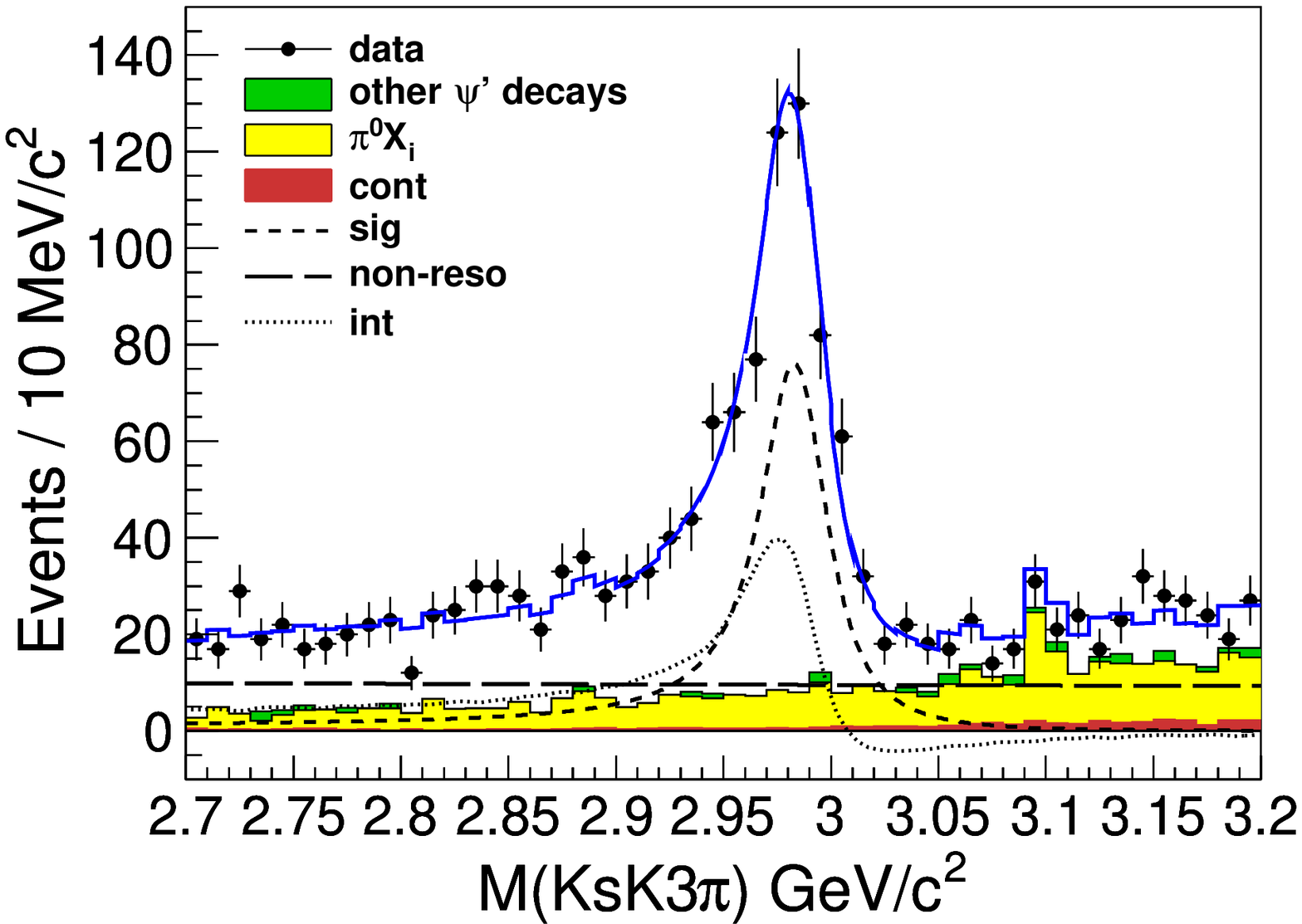}
\includegraphics[width=0.3\columnwidth]{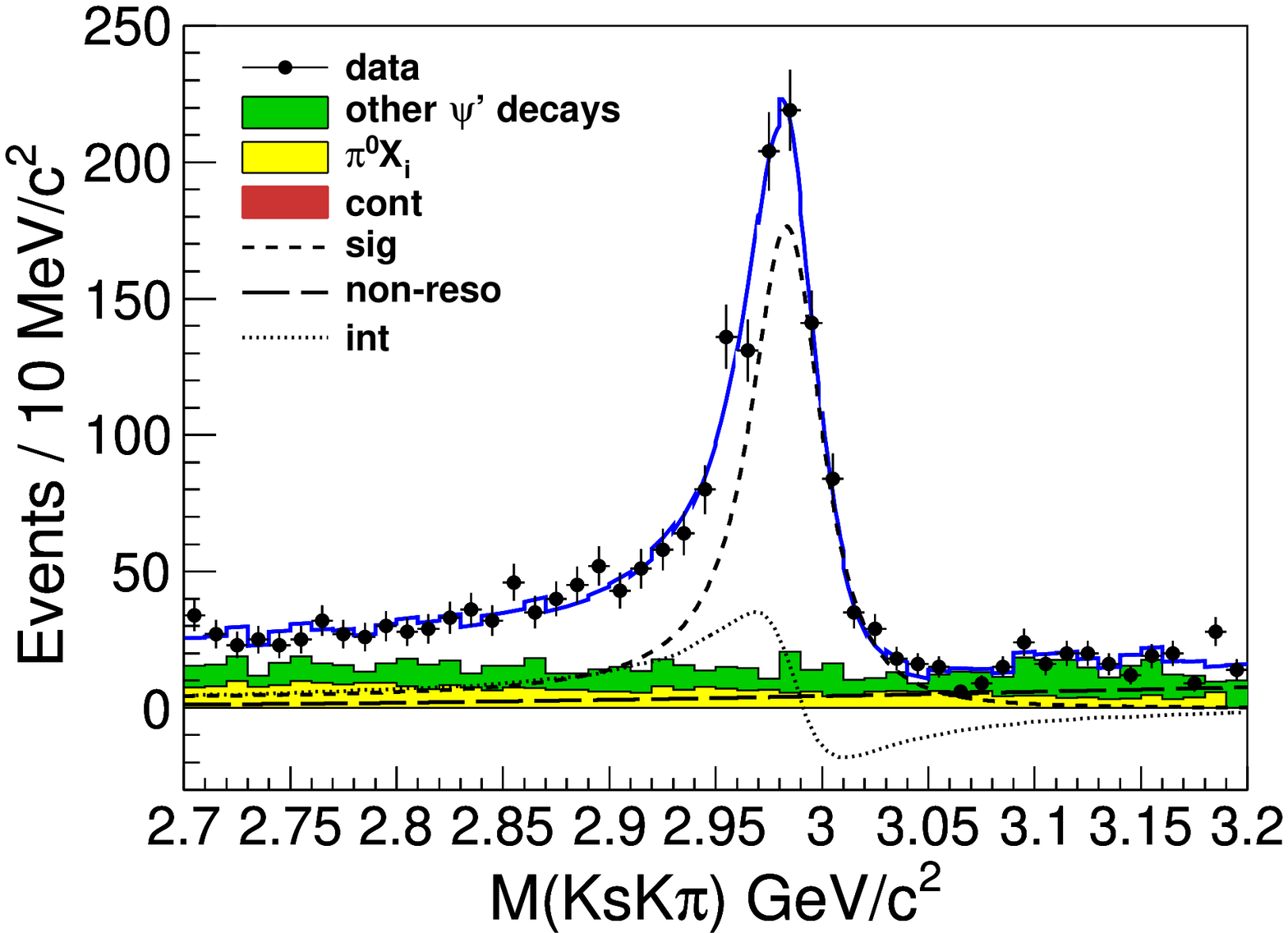}
\includegraphics[width=0.3\columnwidth]{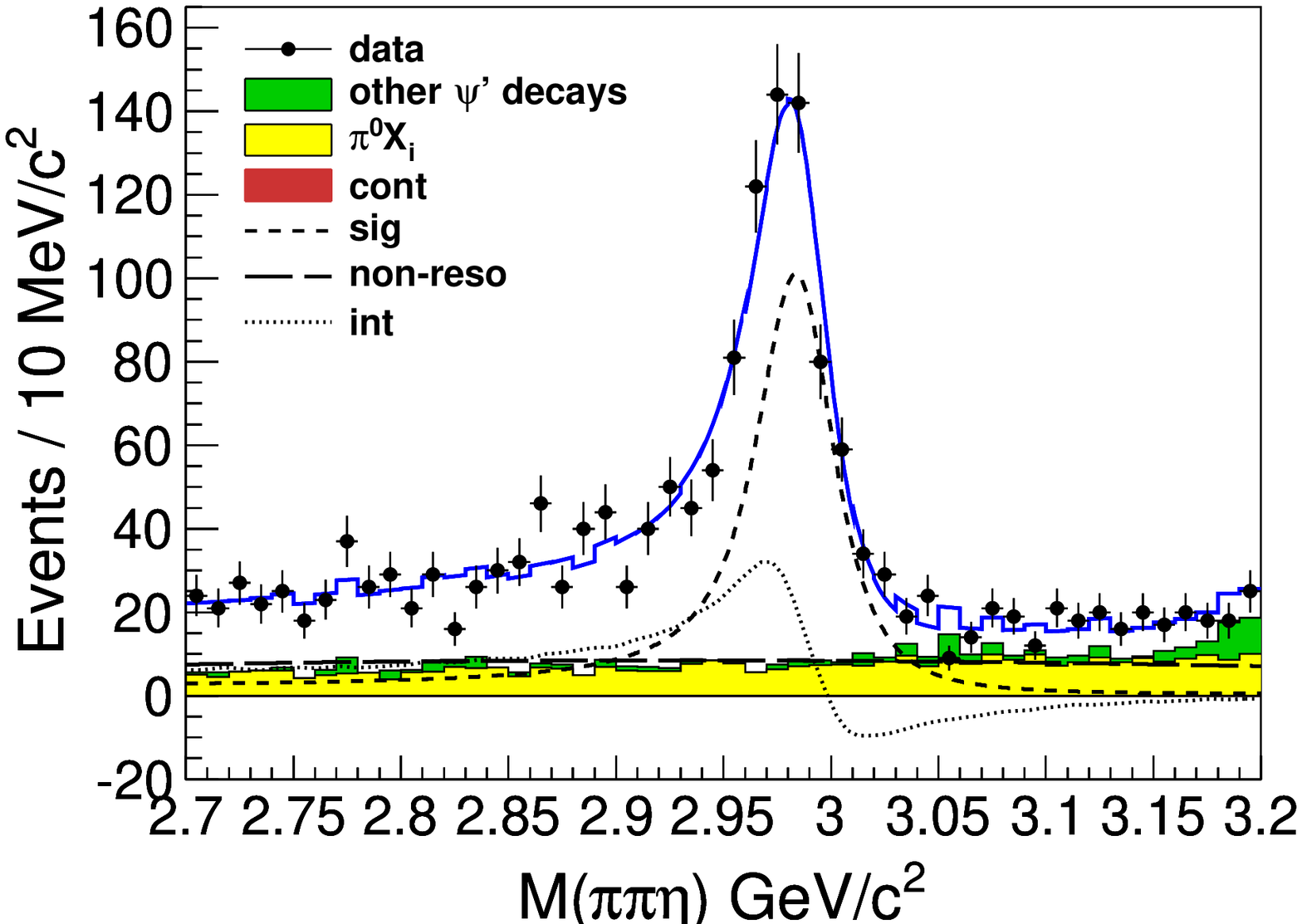}
\includegraphics[width=0.3\columnwidth]{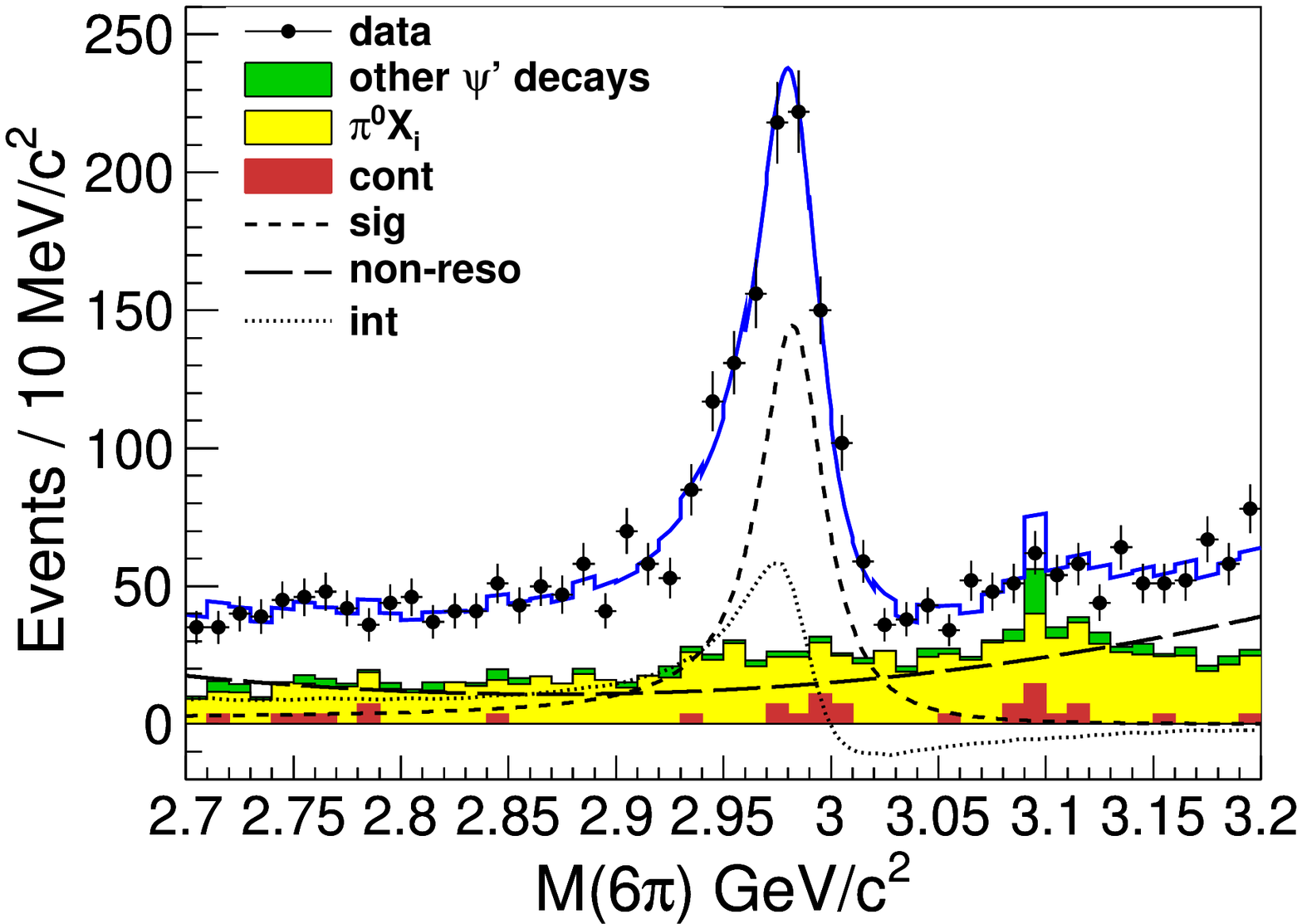}
\caption{The invariant mass distributions for the decays
$K^+K^-\pi^+\pi^-\pi^0$,$K^+K^-\pi^0$,$K_s K^+\pi^-\pi^+\pi^-$,$K_s
K\pi$,$\eta \pi^+\pi^-$ and 3$(\pi^+\pi^-)$, respectively. Solid
curves show the fitting results; the fitting components ($\eta_c$
signal/non-resonance/interference) are shown as
(dashed/long-dashed/dotted) curves. Points with error bar are data,
shaded histograms are (in green/yellow/magenta) for (continuum/other
$\eta_c$ decays/other $\psi(2S)$ decays) backgrounds.}
\label{fig:eta-c-shape}
\end{center}
\end{figure}

Based on the data sample of 106 M $\psi(2S)$ events collected with
the  BESIII detector,  the $\eta_c$ mass and width are measured from
the radiative transition $\psi(2S) \rightarrow \gamma
\eta_c$~\cite{besiiietac}. The $\eta_c$ candidates are reconstructed
from six exclusive decay modes: $K^+K^-\pi^+\pi^-\pi^0$,
$K^+K^-\pi^0$,$K_s K^+\pi^-\pi^+\pi^-$, $K_s K\pi$, $\eta
\pi^+\pi^-$, and 3$(\pi^+\pi^-)$, where $K_s$ is reconstructed in
$\pi^+\pi^-$ mode, $\eta$ and $\pi^0$ from $\gamma \gamma$ final
states. For a hindered M1 transition the matrix element acquires
terms proportional to $E^3_{\gamma}$, which, when combined with the
usual $E^4_{\gamma}$ term accounting for the wave function mismatch
between the radial excited $\psi(2S)$ and the ground-state
$\eta_c(1S)$ transitions, lead to contributions in the radiative
width proportional to $E^7_{\gamma}$. Thus, the $\eta_c$ lineshape
is described by a BW modified by $E_{\gamma}^7$ convoluted with a
resolution function. It is important to point out that the
interference between $\eta_c$ and non-resonance in the signal region
is also considered. The statistical significance of the interference
is 15$\sigma$. This affects the $\eta_c$ resonant parameters
significantly. By assuming all non-resonant events interfere with
the $\eta_c$, the BESIII analysis obtains $\eta_c$ mass and width,
$M = 2984.3 \pm 0.6 \pm 0.6$ MeV/$c^2$ and $\Gamma = 32.0 \pm 1.2
\pm 1.0$ MeV, respectively. Two solutions of relative phase are
found for each decay mode, one represents constructive interference,
the other for destructive. Regardless which solution one takes, the
mass, width of the $\eta_c$ and the overall fit are always
unchanged. The relative phases for constructive interference or
destructive interference from each mode are consistent with each
other within 3$\sigma$, which may suggest a common phase in all the
modes under study. The fit with common phase for each mode shows the
relative phase $\phi = 2.40 \pm 0.07\pm0.08$ rad (constructive) or
$\phi=4.19\pm0.03\pm0.09$ rad (destructive).  The physics that the
interference phase is the same for the six channels is yet to be
understood. Figure~\ref{fig:eta-c-shape} shows the fit results in
the six $\eta_c$ decay modes.

With precise measurement of the $\eta_c$ mass, one can obtain the
hyperfine splitting, $\Delta M_{hf}(1S)_{c\bar{c}} \equiv M(J/\psi)
- M(\eta_c) = 112.6 \pm 0.8$ MeV, which agrees with the quark model
prediction~\cite{kseth}, and will be helpful for understanding the
spin-dependent interactions in hidden charmonium states.

\subsubsection{Observation of $\psi(2S) \rightarrow \gamma
\eta_c(2S)$}

The first radially excited S-wave spin singlet state in the
charmonium system, $\eta_c(2S)$, was observed by the Belle
experiment in the decay process $B^\pm \rightarrow K^\pm
\eta_c(2S)$, $\eta_c(2S) \rightarrow K_s K^\pm
\pi^\mp$~\cite{belle-etac-2s}. It was confirmed by the
CLEO~\cite{cleo-etac-2s} and BABAR~\cite{babar-etac-2s} experiments
in the two-photon fusion process $e^+e^- \rightarrow e^+e^-
(\gamma\gamma), \gamma \gamma \rightarrow \eta_c(2S)\rightarrow
K_sK^\pm\pi^\mp$ and by the BABAR experiment in the
double-charmonium production process $e^+e^- \rightarrow J/\psi
(c\bar{c})$~\cite{babar-etac-2s-double}. The only evidence for
$\eta_c(2S)$ in the $\psi(2S) \rightarrow \gamma \eta_c(2S)$ decay
was from Crystal Ball Collaboration~\cite{crystal-ball-etac-2s} by
looking at the radiative photon spectrum. Recently, the CLEO-c
experiment searched for the $\psi(2S) \rightarrow \gamma \eta_c(2S)$
signal with $\eta_c(2S)$ exclusive decay into 11 modes by using 25.9
M $\psi^\prime$ events, and no evidence found. Product branching
fraction upper limits are determined as a function of
$\Gamma(\eta_c(2S))$ for the 11 individual
modes~\cite{cleo-c-etac-2s}.
\begin{figure}[ht]
\begin{center}
\includegraphics[width=0.3\columnwidth]{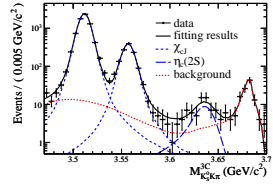}
\caption{Preliminary fitting of the mass spectrum for $\eta_c(2S)
\rightarrow K_s K^\pm \pi^\mp$.} \label{fig:etac-2s}
\end{center}
\end{figure}

The BESIII experiment searched for the M1 transition $\psi(2S)
\rightarrow \gamma \eta_c(2S)$ through the hadronic final states
$K_sK^\pm\pi^\mp$. A bump is observed around 3635 MeV/$c^2$ on the
mass spectrum as shown in Fig.~\ref{fig:etac-2s}. In order to
determine the background and mass resolution using data, the mass
spectrum range is enlarged (3.47 $\sim$ 3.72 GeV/c$^2$) to include
$\chi_{c1}$ and $\chi_{c2}$ events. The resonances $\chi_{c1}$ and
$\chi_{c2}$ are described by the corresponding Monte Carlo (MC)
shape convolved a Gaussian which takes account the small difference
on the mass shift and resolution between data and MC. So the mass
resolution for the $\eta_c(2S)$ in the fitting is fixed to the
linear extrapolation of the mass resolutions from the $\chi_{c1}$
and $\chi_{c2}$ signals in data. The lineshape for $\eta_c(2S)$
produced by such the M1 transition is described by $(E_{\gamma}^3
\times BW(m)\times damping(E_\gamma)) \otimes Gauss(0,\sigma) $
where $m$ is the invariant mass of $K_s K^\pm \pi^\mp$, $E_\gamma =
\frac{m^2_{\psi^\prime}-m^2}{2m^2_{\psi^\prime}}$ is the energy of
the transition photon in the rest frame of $\psi(2S)$,
$damping(E_\gamma)$ is the function to damp the diverging tail
raised by $E^3_\gamma$ and $Gauss(0,\sigma)$ is the Gaussian
function describing the detector resolution. The possible form of
the damping function is somewhat arbitrary, and one suitable
function used by KEDR for a similar process is~\cite{kedr}
$$
\frac{E_0^2}{E_\gamma E_0 + (E_\gamma E_0 - E_0)^2}
$$
where $E_0 =\frac{m^2_{\psi(2S)}-m_{\eta_c(2S)}^2}{2m^2_{\psi(2S)}}$
is the peaking energy of the transition photon. In the fit, the
width of $\eta_c(2S)$ is fixed to PDG value. From the fit to the
data, a signal with a statistical significance of 6.5 standard
deviation is observed which is the first observation of the M1
transition $\psi(2S) \rightarrow \gamma \eta_c(2S)$. The measured
mass for $\eta_c(2S)$ is $3638.5 \pm 2.3 \pm 1.0$ MeV/$c^2$. The
measured branching ratio is $\mathcal{BR}(\psi(2S) \rightarrow
\gamma \eta_c(2S))\times \mathcal{BR}(\eta_c(2S) \rightarrow
K_sK^\pm\pi^\mp) = (2.98 \pm 0.57\pm 0.48) \times 10^{-6}$. Together
with the BABAR result $\mathcal{BR}(\eta_c(2S)\rightarrow
K\bar{K}\pi) = (1.9 \pm 0.4 \pm1.1)\%$~\cite{babar-gg-1}, the M1
transition rate for $\psi(2S) \rightarrow \gamma \eta_c(2S)$ is
derived as $\mathcal{BR}(\psi(2S) \rightarrow \gamma \eta_c(2S))=
(4.7 \pm 0.9 \pm 3.0)\times 10^{-4}$, which is in agreement with the
prediction of potential model calculations~\cite{etac2smodel}.

Recently, the Belle experiment measured $\eta_c(1S)$ and
$\eta_c(2S)$ resonant parameters in the decay of $B^\pm \rightarrow
K^\pm (K_sK\pi)^0$ by considering the interference between
$\eta_c(1S)$/$\eta_c(2S)$ decay and non-resonant in the $B^\pm$
decay~\cite{belle-etac-update}. Meanwhile, the BABAR experiment also
updated the analysis of $e^+e^- \rightarrow e^+e^-(\gamma\gamma),
\gamma\gamma \rightarrow \eta_c(1S)/\eta_c(2S)\rightarrow
(K_sK\pi)^0$ and $K^+K^-\pi^+\pi^-\pi^0$
modes~\cite{babar-etac-update}. Table~\ref{table1} shows the summary
of the typical production processes ($\psi(2S)$ radiative decays,
$\gamma\gamma$ fusion and $B$ decays) for $\eta_c(1S)/\eta_c(2S)$
and corresponding resonant parameter measurements. These results
indicate that the $\eta_c(1S)$ parameters agree well from different
production processes.
\begin{table}[htb]
  \begin{center}
  \caption{Comparison of the mass and width for $\eta_c(1S)/\eta_c(2S)$ in different production processes,
    $\psi^\prime\rightarrow \gamma \eta_c(1S)/\eta_c(2S)$, $B^\pm \rightarrow K^\pm \eta_c(1S)/\eta_c(2S)$ and
     $\gamma \gamma$ fusion, from different experiments. The PDG values are only world average from earlier results.
     . For the time being, the most precise measurements for $\eta_c(1S)$ resonance are from BESIII, while these for
     $\eta_c(2S)$ resonant parameters are from BABAR in $\gamma\gamma$ fusion.}
    \label{table1}
{\scriptsize
    \begin{tabular}{lcccc}\hline
              & BESIII & Belle~\cite{belle-etac-update} & BABAR~\cite{babar-etac-update} & PDG 2010~\cite{pdg2010}              \\
              & $\psi^\prime \rightarrow \gamma \eta_c/\eta_c(2S)$ &
              $B$ decays & $\gamma\gamma$ fusion &   \\
      \hline
      $M(\eta_c(1S))$ MeV/$c^2$   &$2984.3\pm0.6\pm0.6$ & $2985.4\pm1.5^{+0.2}_{-2.0}$ &  $2982.2\pm0.4\pm1.4$ & $2980.3\pm1.2$  \\
      $\Gamma(\eta_c(1S))$ MeV & $32.0\pm1.2\pm1.0$ & $35.1\pm3.1^{+1.0}_{-1.6}$ & $32.1\pm1.1\pm1.3$& $28.6\pm2.2$  \\
$M(\eta_c(2S))$ MeV/$c^2$ & $3638.5\pm2.3\pm1.0$ &
$3636.1^{+3.9+0.5}_{-1.5-2.0}$ & $3638.5\pm1.5\pm0.8$ & $3637\pm4$
\\
$\Gamma(\eta_c(2S))$ MeV & $12$ (fixed) &
$6.6^{+8.4+2.6}_{-5.1-0.9}$ & $13.4\pm4.6\pm3.2$ & $14\pm7$ \\
\hline
    \end{tabular}}
  \end{center}
\end{table}

\subsection{New results on $h_c$ from BESIII }

\begin{figure}[ht]
\begin{center}
\includegraphics[width=0.3\columnwidth]{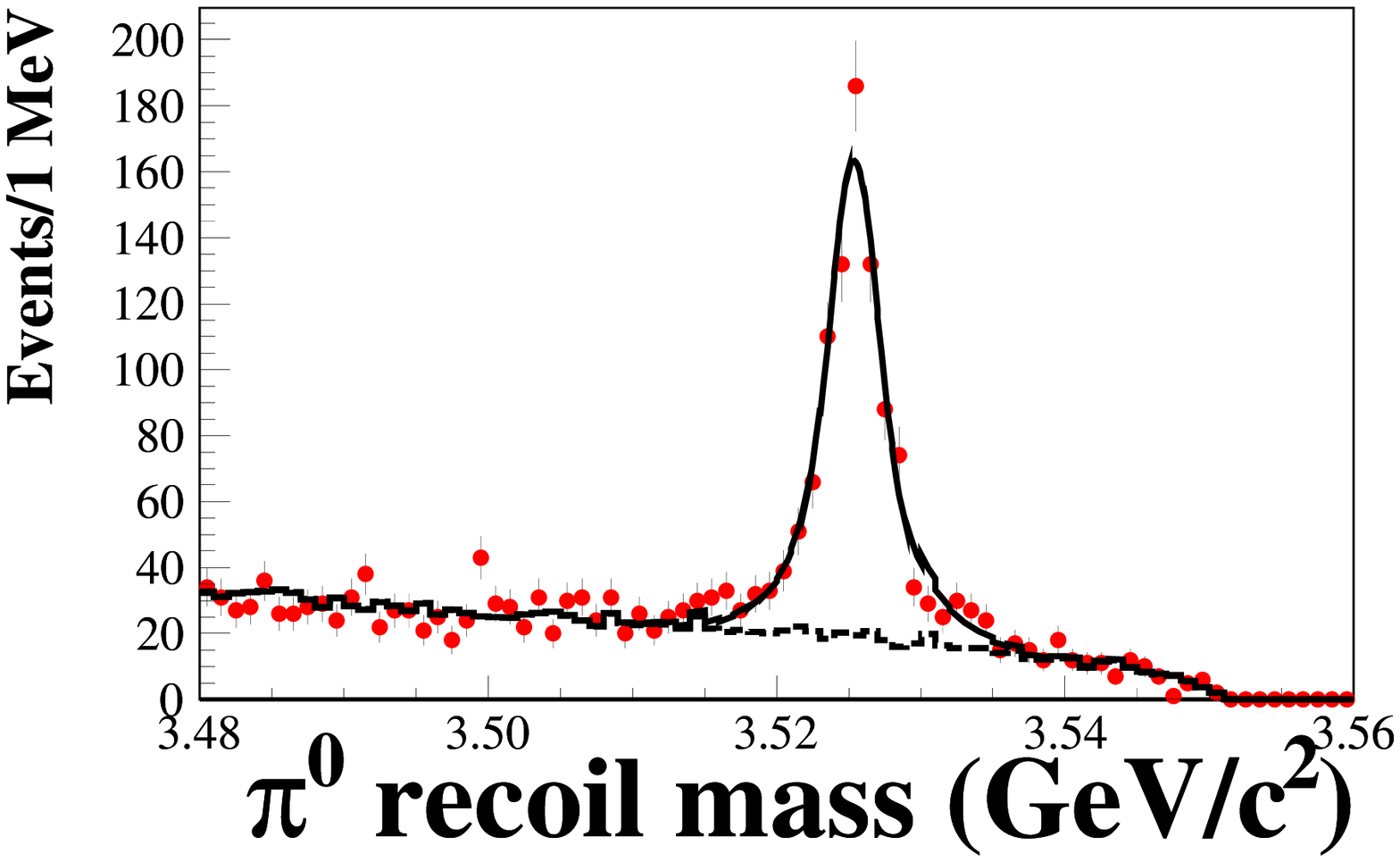}
\includegraphics[width=0.3\columnwidth]{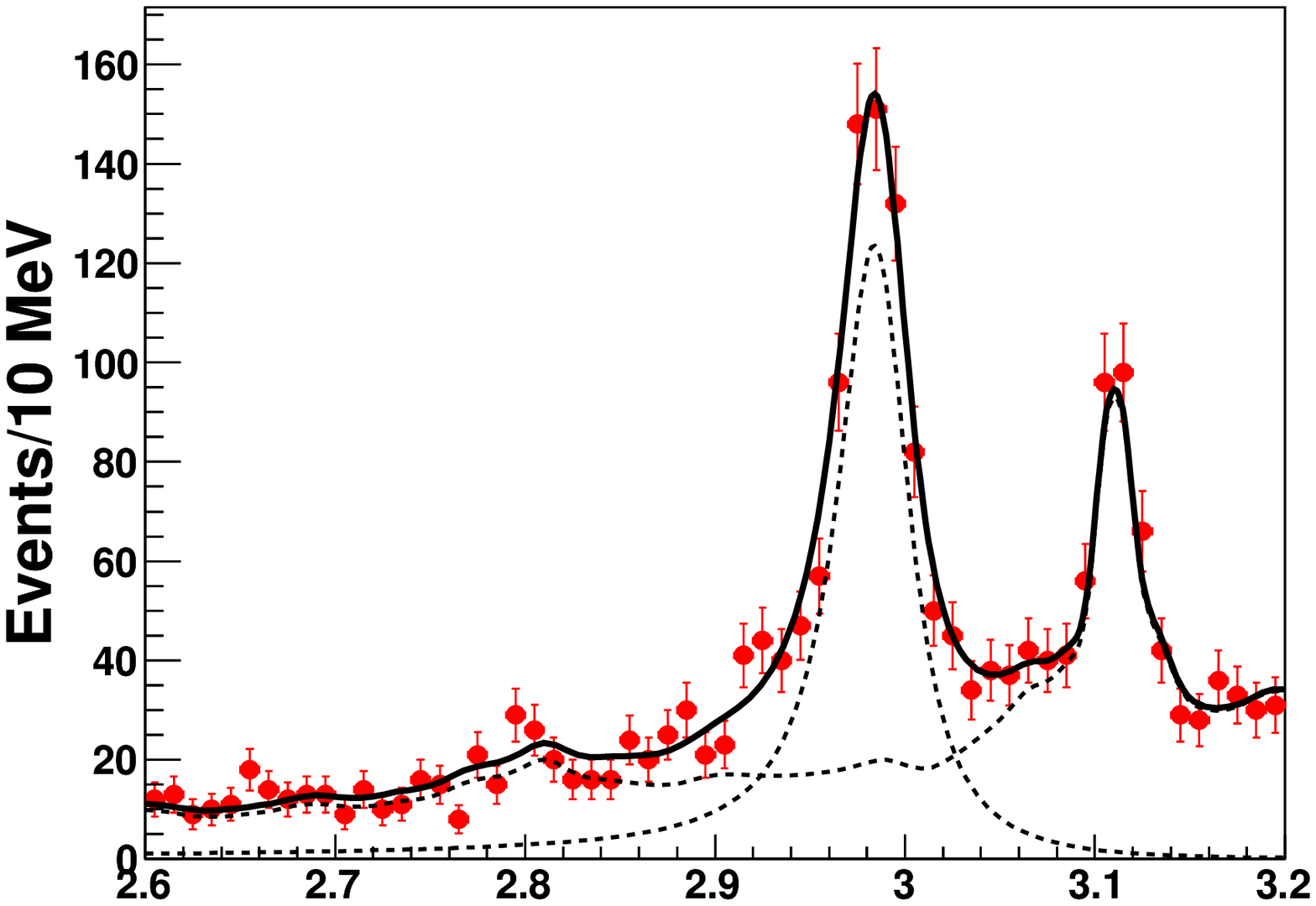}
\caption{Left: The $\pi^0$ recoiling mass for the sum of 16 $\eta_c$
decay modes and fitting results. Right: The invariant mass
distributions for the sum of 16 $\eta_c$ decay modes, and fitting
results. All are preliminary results. } \label{fig:hc-etac}
\end{center}
\end{figure}

The BESIII Collaboration reported the results on the production and
decay of the $h_c$ using 106M of $\psi(2S)$ decay events in
2010~\cite{hc_paper}, where they studied the distributions of mass
recoiling against a detected $\pi^0$ to measure $\psi(2S)
\rightarrow \pi^0 h_c$ both inclusively (E1-untagged) and in events
tagged as $h_c \rightarrow \gamma \eta_c$ (E1-tagged) by detection
of the E1 transition photon. In 2011, 16 specific decay modes of
$\eta_c$ are used to reconstruct $\eta_c$ candidates in the decay
mode of $h_c\rightarrow \gamma \eta_c$. Figure~\ref{fig:hc-etac}
(left) shows the $\pi^0$ recoiling mass for the sum of the 16
$\eta_c$ decay modes.  Fits to the 16 $\pi^0$ recoil-mass spectra
are performed simultaneously that yields $M(h_c)= 3525.31\pm 0.11
\pm0.15$ MeV/$c^2$ and $\Gamma(h_c)= 0.70\pm0.28 \pm0.25$ MeV/$c^2$,
respectively. These preliminary results are consistent with the
previous BESIII inclusive results and CLEO-c exclusive results.

The centroid of the $^3P_J$ states
($\chi_{c0}$,$\chi_{c1}$,$\chi_{c2}$) is known to be $\langle
M(^3P_J)\rangle  = [5M(^3P_2) + 3M(^3P_1) + M(^3P_0)] = 3525.30 \pm
0.04$ MeV~\cite{pdg2010}. If the $^3P_J$ states centroid mass
$\langle M(^3P_J) \rangle$ is identified as the mass of $M(^3P)$,
then BESIII observes the hyperfine splitting as $\Delta
M_{hf}(1P)_{cc} = -0.01 \pm 0.11 \pm 0.14 $ MeV which agrees with
zero.

The BESIII Collaboration also looked at the $\eta_c(1S)$ mass
distributions in the exclusive $h_c \rightarrow \gamma \eta_c(1S)$
decay modes. Figure~\ref{fig:hc-etac} (right) shows the distribution
of invariant mass distribution from exclusive hadronic decays of
$\eta_c(1S)$, the shape of the $\eta_c$ is symmetric in $h_c(1P)
\rightarrow \gamma \eta_c(1S)$ radiative transition and no
distortion observed as in $J/\psi$/$\psi^\prime$ radiative
decays~\cite{cleo-c-new,besiiietac}. A maximum likelihood fit is
perform to extract the $\eta_c(1S)$ parameters. In the fit the
signal shape is described by a BW (corrected by $E^3_\gamma$)
convoluted with a Gaussian resolution function. The preliminary
results for the $\eta_c(1S)$ parameters are $M=2983.6\pm 1.1$
MeV/$c^2$ and $\Gamma = 36.6 \pm 3.15$ MeV/$c^2$, where the errors
are statistical error only. The results are in agreement with those
from $\psi(2S) \rightarrow \gamma \eta_c(1S)$ analysis.

\subsection{Determination of $\psi(3770)$ parameters at KEDR }

A measurement of the $\psi(3770)$ meson parameters has been reported
by the KEDR experiment based on data samples during the scans of the
center-of-mass energy range from 3.67 to 3.92 GeV at the VEPP-4M
$e^+e^-$ collider in 2004 and 2006~\cite{kedr2011}. The observed
multihadron cross sections were fitted as a function of the
center-of-mass energy using some assumptions about the behaviour of
then non-resonant form factor. The data points corrected for the
detector efficiency together with several fitting curves are shown
in Fig.~\cite{fig:kedr} (left). Unlike Ref.~\cite{besii3770}, KEDR
experiment did not observe any shape anomaly. The interference of
resonant and nonresonant production essential in the near-threshold
region has been taken into account as suggested in Ref.~\cite{yang}.
The analysis assumes the domination of $\psi(2S)$ in the
non-$\psi(3770)$ part of the $D$-meson form factor. It is also worth
noting that correct accounting of the resonance-continuum
interference can help in solving the $\psi(3770)$ non-$D\bar{D}$
puzzle~\cite{liyang}.
\begin{figure}[ht]
\begin{center}
\includegraphics[width=0.3\columnwidth]{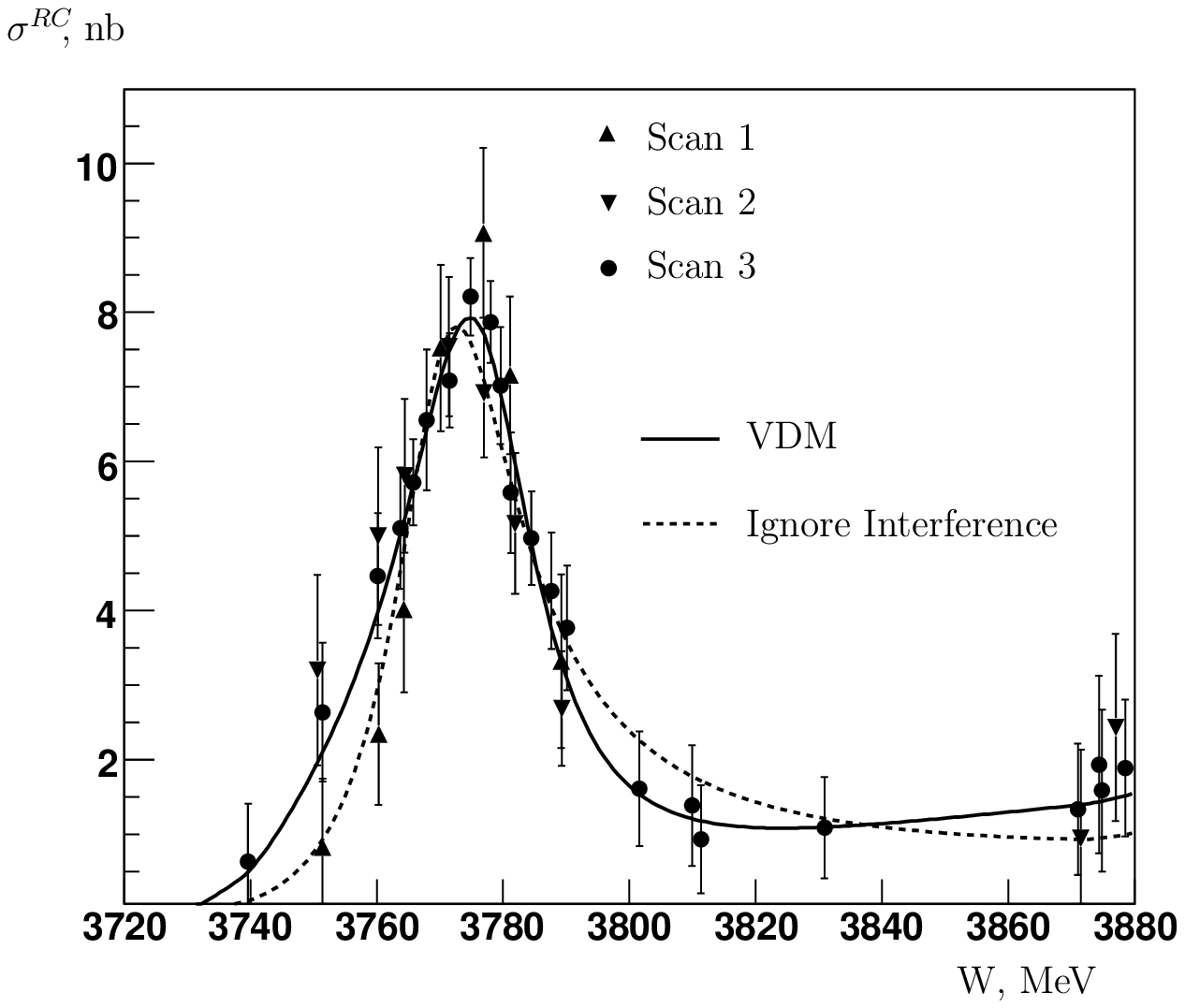}
\includegraphics[width=0.3\columnwidth]{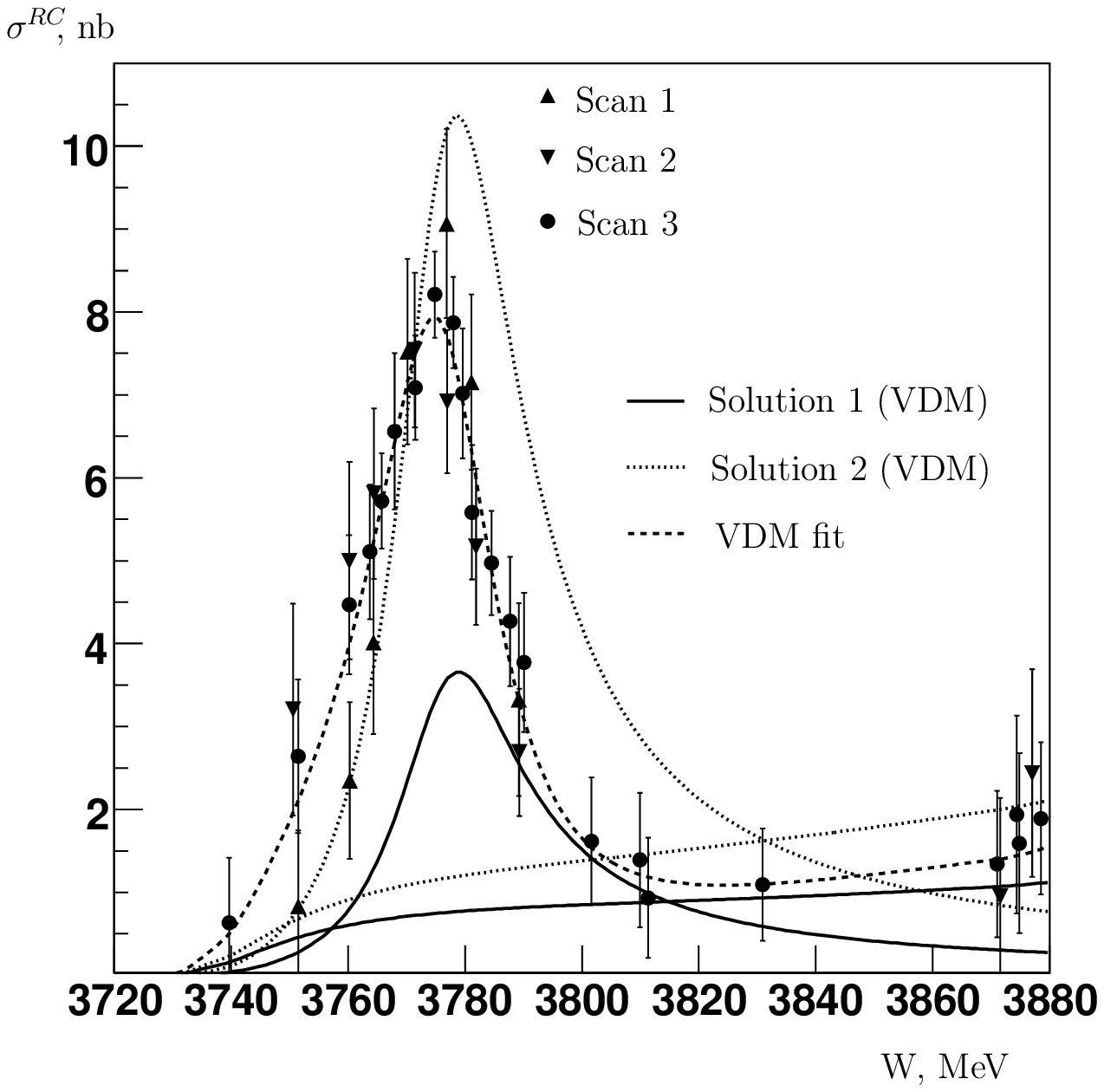}
\caption{Left: Cross section of  $e^+e^-\rightarrow hadrons$  vs.
c.m. energy in the vicinity of $\psi(3770)$ with the light quark,
$\tau$ and QED backgrounds subtracted. The curves are the result of
a simultaneous fit for different models. Right: The same plot as
left one, solid and short-dashed curves correspond to two VDM
solutions. Resonant and  non-resonant parts are presented
separately. }
  \label{fig:kedr}
\end{center}
\end{figure}

It is known that for two interfering resonances the ambiguity can
appear in the resonance amplitudes and the interference
phase~\cite{twosolution}. Figure~\ref{fig:kedr} (right) shows the
results of the two solutions corresponding to vector dominance
models (VDM).  The KEDR experiment presented the results on the mass
ad width of $\psi(3770)$ are $M=3779.1^{+1.8}_{-1.7} \pm
0.6^{+0.2}_{-0.3}$ MeV/$c^2$ and $\Gamma = 25.2^{+4.6}_{-4.1} \pm
0.5^{+0.5}_{-0.2}$ MeV, where average values of the mass and total
width of the two VDM solutions are used and the third error arises
from the model dependence. The results agree with those from
BABAR~\cite{babar3770} also taking into account interference and
disagree with all results obtained ignoring this effect including
that by BESII~\cite{besii3770-p}. The KEDR also obtained two
possible solutions for the $\psi(3770)$ electron partial width
\begin{enumerate}
\item $\Gamma_{ee} = 147^{+97}_{-62} \pm
13^{+11}_{-10}$ eV,
\item $\Gamma_{ee} = 415^{+59}_{-58} \pm
38^{+160}_{-10}$ eV.
\end{enumerate}
Most of the potential models strongly support the first solution and
can barely tolerate the second one~\cite{pmodel1,pmodel2,pmodel3}.

\subsection{News on charmonium-like states}

\subsubsection{$X(3872)$}

The $X(3872)$ resonance is the forerunner of the new charmonium
family. It was discovered by Belle~\cite{first-belle} in the $B
\rightarrow KX$ decay, with $X\rightarrow J/\psi \pi^+\pi^-$, and
confirmed by BABAR~\cite{babar-3872,babar-3872-1},
CDF~\cite{cdf-3872} and D0~\cite{d03872},  and now by
CMS~\cite{cms-3872-1,cms-3872-2} and LHCb~\cite{lhcb-3872}. The
quantum number $J^{PC}$ of the $X(3872)$ state is not yet fully
established. A study of the angular decay distributions performed by
the CDF experiment gave nonnegligible probabilities only for
$1^{++}$ and $2^{-+}$~\cite{cdf-3872}, where $C=+$ is also confirmed
by the observation of the $X\rightarrow J/\psi\gamma$
decay~\cite{belle-3872-2}.  Table~\ref{tab:sum3872} summarizes the
measurements of the $X(3872)$ mass. The new world average is $M=
3871.67\pm0.17$. The close proximity of the world-averaged mass to
the $D^{*0}\bar{D}^0$ ($m_{D^{*0}} + m_{\bar{D}^0}=3871.79 \pm 0.30$
MeV/$c^2$) mass threshold has engendered speculation that the
$X(3872)$ might be a loosely bound $D^{*0}\bar{D}^0$ molecule or a
tetraquark state~\cite{ddbar3872}. Despite a large experimental
effort, the nature of this new state is still uncertain and several
models have been proposed to describe it.
\begin{table}[!t]
  \caption{Comparison of the $X(3872)$ mass measurements from different experiments, and world average.}
  \medskip
  \label{tab:sum3872}
\centering { \footnotesize
  \begin{tabular}{lc} \hline \hline
 Experiment &
               $X(3872)$ mass MeV/$c^2$
\\ \hline
           CDF &
           $3871.61\pm 0.16 \pm0.19$
 \\
           BABAR ($B^\pm$) &
           $ 3871.4\pm0.6\pm0.1$
 \\
            BABAR ($B^0$)&
           $3868.7\pm1.5\pm 0.4$
 \\
           D0 &
           $3871.8\pm 3.1\pm3.0$
 \\
            Belle (full dataset)             &
           $3871.84\pm0.27\pm0.19$
 \\
           LHCb &
           $3871.96\pm0.46\pm0.10$
\\\hline
         World Average  &
          $3871.67\pm0.17$
\\
\hline \hline
\end{tabular}}
\end{table}

With full dataset (711 fb$^{-1}$) collected by the Belle detector,
the difference in masses of the $X(3872)$ states produced in $B^+
\rightarrow  K^+ \pi^+\pi^- J/\psi$ and $B^0 \rightarrow K^0
\pi^+\pi^- J/\psi$ decays is determined to be~\cite{belle-3872-1}:
$\Delta M_{X(3872)} = (-0.69 \pm0.97 \pm0.19)$ MeV/$c^2$, which is
consistent with zero and disagrees with theoretical predictions
based on a diquark model for the $X(3872)$~\cite{diquark}. We
conclude from this that the same particle is produced in the two
processes and use a fit to the combined neutral and charged $B$
meson data samples to determine: $M_{X(3872)} = (3871.84 \pm 0.27
\pm0.19 )$ MeV/$c^2$. This result agrees with the current
world-average value of $3871.67\pm0.17$ MeV/$c^2$ as listed in
Table~\ref{tab:sum3872}. The width of the $X(3872)$ signal peak is
consistent with the experimental mass resolution and we set a 90\%
CL limit on its natural width of $\Gamma_{X(3872)} < 1.2$ MeV,
improving on the previous limit of 2.3 MeV. The details of the
analysis can be found in~\cite{belle-3872-1}.  The production ratio
of $X(3872)$ in the $B^0$ and $B^+$ meson decays are measured to be
:
\begin{equation}
R(X) = \frac {{\mathcal B}(B^0\rightarrow K^0 X(3872))} {{\mathcal
B}(B^+\rightarrow K^+ X(3872))} =   0.50\pm 0.14 \pm 0.04,
 \label{eq:x3872}
\end{equation}
This value is above the range preferred by some molecular models for
the $X(3872)$: $0.06\le R(X) \le 0.29$~\cite{swanson_pr}. The BaBar
result for this ratio is $R(X) = 0.41 \pm 0.24 \pm
0.05$~\cite{babar-3872-1}. Both BABAR~\cite{babar3872charge} and
Belle~\cite{belle-3872-1} experiments had searched the charged
partner of the $X(3872)$ decaying into $\pi^+\pi^0J/\psi$ by using
$B \rightarrow K \pi^+\pi^0 J/\psi$ decay, and shows no evidence for
a charged partner to the $X(3872)$ decaying as $X^+ \rightarrow
\rho^+ J/\psi$. The current data favor isospin $I = 0$. However, the
close proximity of the $D^{*0}\bar{D}^0$ threshold may induce large
isospin violations, as pointed out by
N.~A.~Tornqvist~\cite{Tornqvist}.  Unfortunately, in Belle
experiment, an attempt to discriminate between $J^{PC} = 1^{++}$ and
$2^{-+}$ with an angular analysis of $X(3872) \rightarrow \pi^+\pi^-
J/\psi$ didn't brought to any conclusion, due to the lack of
statistics~\cite{belle-3872-1}.

Radiative decays of the $X(3872)$ are important in understanding its
nature. The decay of $X(3872)\rightarrow J/\psi \gamma$ established
its charge parity to be +1. Both BaBar~\cite{babar3872-g} and
Belle~\cite{belle-3872-2} searched for the the $X(3872)\rightarrow
J/\psi \gamma$ and $X(3872)\rightarrow \psi(2S) \gamma$
 decays, that are predicted to dominate for a molecule.
The BABAR Collaboration~\cite{babar3872-g} show that
$\mathcal{BR}(X(3872)\rightarrow \psi(2S) \gamma)$ is almost three
times that of $\mathcal{BR}(X(3872)\rightarrow J/\psi \gamma)$,
while, the Belle Collaboration find no evidence for
$X(3872)\rightarrow \psi(2S) \gamma$, and determine the ratio to be
: $R \equiv
\frac{\mathcal{B}(X(3872)\to\psi'\gamma)}{\mathcal{B}(X(3872)\to
J/\psi\gamma)}  < 2.1$ (at 90\% C.L.)~\cite{belle-3872-2}. The
possibility of a $\psi(2S) \gamma$ dominance seems to be excluded.
The $X(3872)$ state may not have a large  $c\bar{c}$ admixture  with
a $D^{*0}\overline{D}{}^0$ molecular component as was expected on
the basis of  the BABAR result~\cite{babar3872-g}.

\begin{figure}[ht]
\begin{center}
\includegraphics[width=0.32\columnwidth]{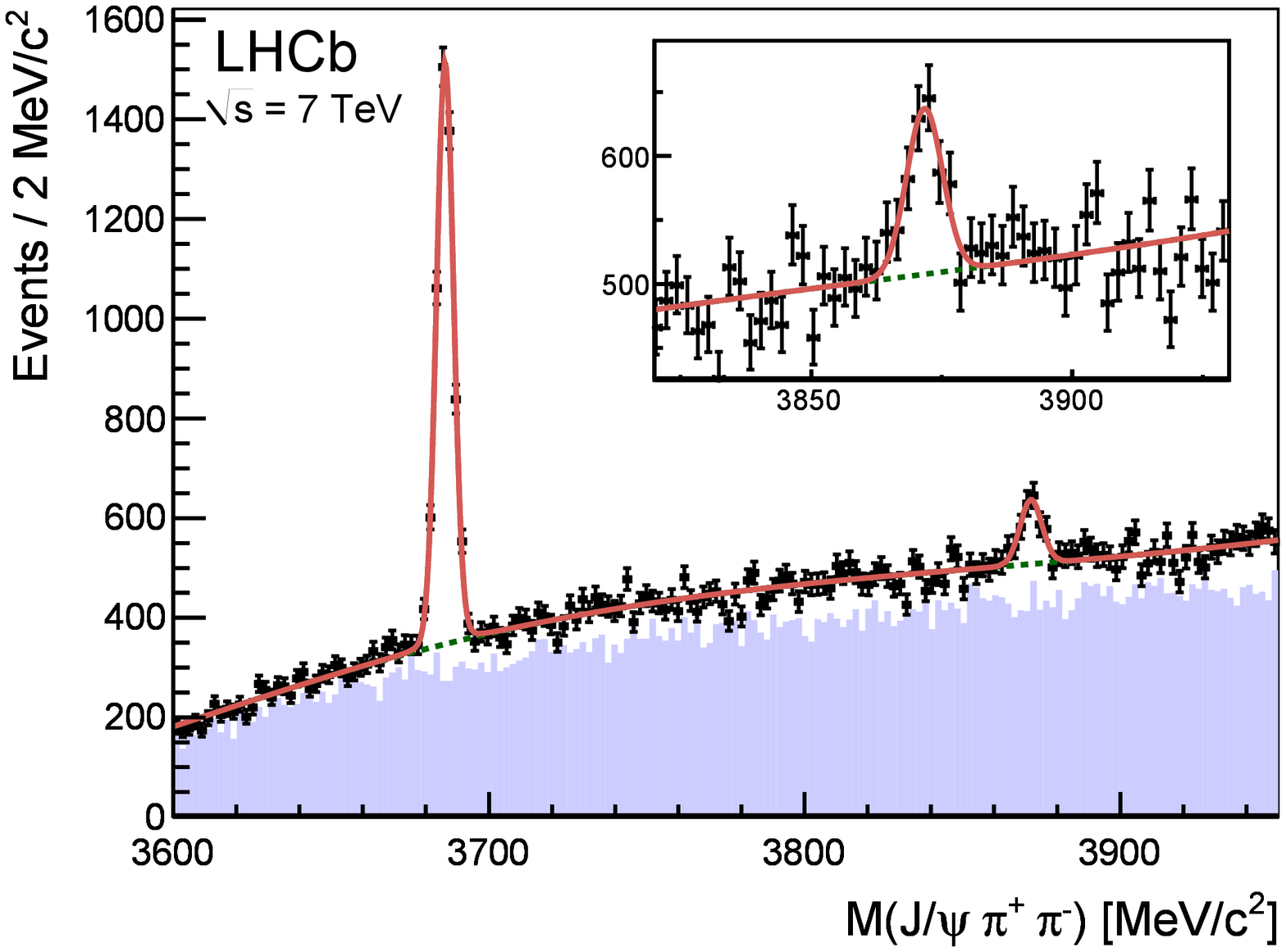}
\includegraphics[width=0.32\columnwidth]{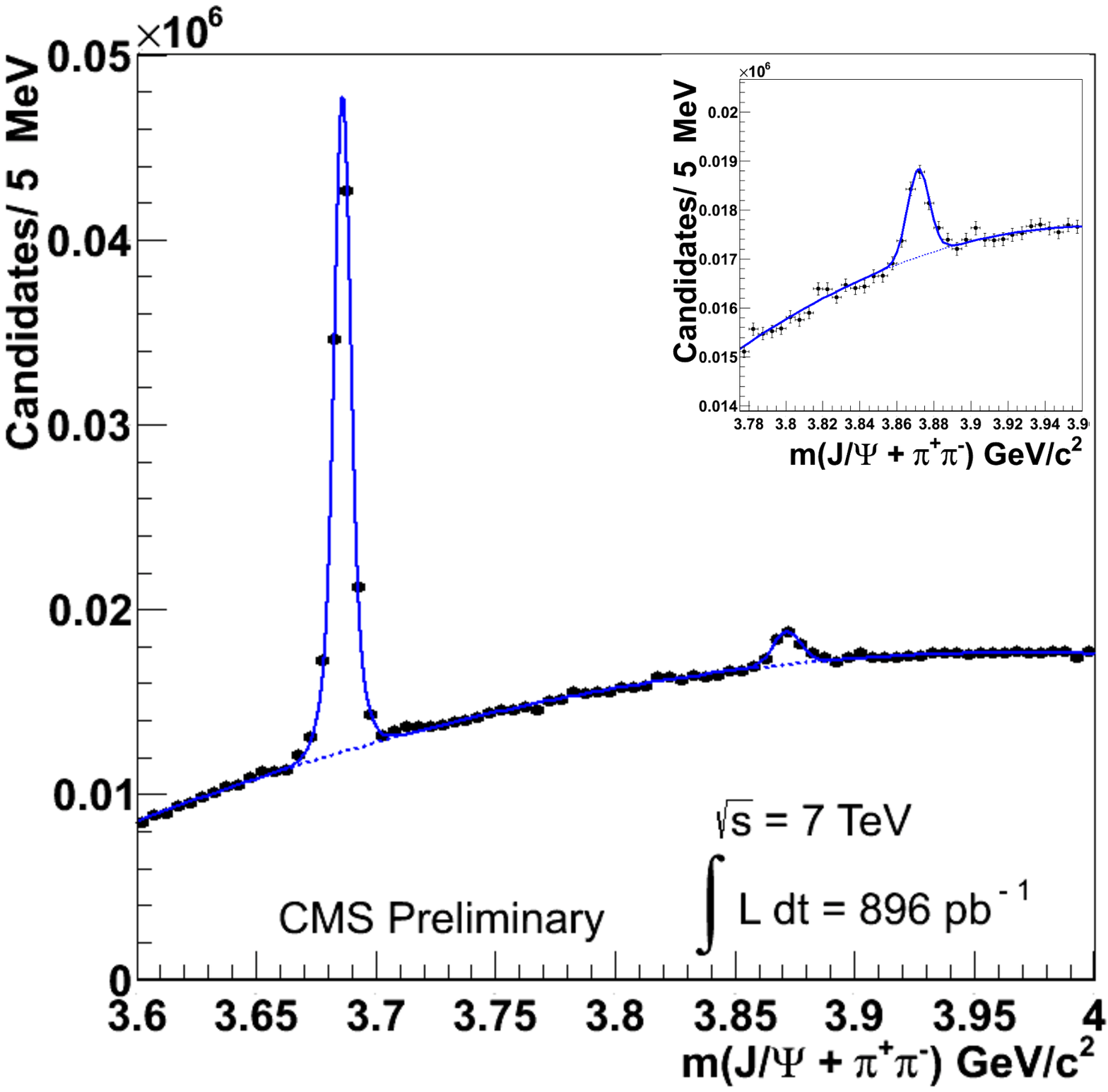}
\caption{Left from LHCb:  Invariant mass distribution of
$J/\psi\pi^+\pi^-$ (black points with statistical error bars) and
same-sign $J/\psi \pi^\pm\pi^\pm$  (blue filled histogram)
candidates. The solid red curve is the result of the fit. Right from
CMS: Invariant mass distribution of $J/\psi\pi^+\pi^-$ from. The
inset shows a zoom of the $X(3872)$ region.}
  \label{fig:lhcb3872}
\end{center}
\end{figure}

First observation of the $X(3872)$ state at the LHC were reported by
both CMS~\cite{cms-3872-1,cms-3872-2} and LHCb~\cite{lhcb-3872}
experiments in the decay of $J/\psi \pi^+\pi^-$, using data
collected in 2010. With an integrated luminosity of 34.7 pb$^{-1}$
collected by the LHCb experiment, the production of the $X(3872)$
particle is observed in pp collisions at $\sqrt{s} = 7$ TeV.
Figure~\ref{fig:lhcb3872} (left) shows the results. The masses of
both the $X(3872)$ and $\psi(2S)$ are measured to be : $m_{X(3872)}
= 3871.95 \pm 0.48 \pm 0.12$ MeV/$c^2$ and $m_{\psi(2S)} =
3686.12\pm 0.06  \pm 0.10$ MeV/$c^2$. The CMS Collaboration
performed a measurement of the $X(3872)$ with data set collected in
2010 and 2011, and found about 5 3000 candidates in 896 pb$^{-1}$,
as shown in Fig.~\ref{fig:lhcb3872} (right).

\subsubsection{$1^{--}$ family}

In the family of the charmonium-like states, the $1^{--}$ states are
among the most easiest to be found at $e^+e^-$ collider, since they
can be produced through the ISR mechanism. The $Y(4260)$ is the
first one found by BABAR Collaboration~\cite{babar4260}. Now four
states were discovered at the
$B$-factories~\cite{babar4260,belle4260,babar4260-1,belle4260-1}:
the $Y(4008)$ and the $Y(4260)$ decaying to $J/\psi \pi^+\pi^-$, the
$Y(4350)$ and the $Y(4660)$ decaying to $\psi(2S) \pi^+\pi^-$.  The
$Y(4260)$ has been also confirmed by the CLEO-c
experiment~\cite{cleo4260,cleo4260-1}  that could extensively
produce it by running at open charm energy region. This allowed to
observe also the $Y\rightarrow J/\psi \pi^0\pi^0$ and the $Y
\rightarrow J/\psi K^+K^-$ modes. The Belle experiment performs an
analysis to compare the $Y(4260) \rightarrow J/\psi \pi^+\pi^-$ and
the $Y(4260)\rightarrow J/\psi \pi^0\pi^0$ rates. According to the
isospin symmetry, the second one should be half of the rate of the
first one. The cross-sections of $e^+e^- \rightarrow J/\psi
\pi^0\pi^0$ as function of mass have been measured, and by fitting
to the cross-sections data, they obtain $\Gamma_{ee}\times
\mathcal{BR}(Y(4260)\rightarrow J/\psi
\pi^0\pi^0)=(3.19^{+1.82+0.64}_{-1.53-0.35})$ eV, which is
consistent with the isospin expectation by comparing to the measured
$\Gamma_{ee}\times \mathcal{BR}(Y(4260)\rightarrow J/\psi
\pi^+\pi^-)$~\cite{pdg2010}.

\subsubsection{$Y(4140)$ at LHCb }

 The CDF experiment reported a 3.8$\sigma$ evidence for the
 $Y(4140)$ state in the $B \rightarrow K Y (Y\rightarrow J/\psi \phi)$
 decay using $p\bar{p}$ data collected at the Tevatron
 ($\sqrt{s}=1.96$ TeV)~\cite{cdf4140}. A preliminary update of the CDF analysis
 with 6.0 fb$^{-1}$ leads to observation of $Y(4140)$ with
 with a statistical significance of more than 5$\sigma$~\cite{cdf4140-1}. The mass
 and width were measured to be $4143^{+2.9}_{-3.0}\pm0.6$ MeV and
 $15^{+10.4}_{-6.1}\pm 2.5$ MeV/$c^2$. The relative decay rate was
 determined to be $\mathcal{BR}(B^+ \rightarrow
 YK^+)\times\mathcal{BR}(Y\rightarrow J/\psi
 \phi)/\mathcal{BR}(B^+\rightarrow J/\psi \phi K^+)=0.149\pm
 0.039\pm 0.024$.  However, the LHCb Collaboration perfrom the most
 sensitive search for the narrow $Y(4140)\rightarrow J/\psi \phi$
 state in $B^+ \rightarrow J/\psi \phi$ decays by using 0.37
 fb$^{-1}$ data. They did not confirm the existence of such a state.
 An upper limit on the $\mathcal{BR}(B^+ \rightarrow
 YK^+)\times\mathcal{BR}(Y\rightarrow J/\psi
 \phi)/\mathcal{BR}(B^+\rightarrow J/\psi \phi K^+)<0.07$ at 90\%
 C.L. is set~\cite{lhcb4140}. The result disagrees at the 2.4$\sigma$ level with the
 CDF measurement.

\section{Bottomonium-like states }

The Belle experiment has collected a large sample of $e^+e^-$
collision at the center-of-mass energy near the $\Upsilon(5S)$
resonance, which lies above the $B_s\bar{B}_s$ production threshold.
Many unexpected non-$B_s\bar{B}_s$ decays of $\Upsilon(5S)$ have
been observed. In particular, anomalously large rates for dipion
transitions $\Upsilon(5S) \rightarrow \Upsilon(nS) \pi^+\pi^-$
($n=1$, 2,3) have been observed~\cite{belle-y-decay}. Assuming these
signals are attributed entirely to the $\Upsilon(5S)$ decays, the
measured partial decay widths $\Gamma (\Upsilon(5S) \rightarrow
\Upsilon(nS) \pi^+\pi^-) \sim 0.5$ MeV are about two order of
magnitude larger than typical widths for dipion transitions of
$\Upsilon(2S)$, $\Upsilon(3S)$ and $\Upsilon(4S)$.

Recently the CLEO-c experiment observed the process $e^+e^-
\rightarrow h_c(1P) \pi^+\pi^-$ at a rate comparable to the process
$e^+e^-\rightarrow J/\psi \pi^+\pi^-$ by using data sample taken
near $\sqrt{s}=4170$ MeV and found evidence of an even high
transition rate at the $Y(4260)$ energy~\cite{cleoc4260}. This
implies that the $h_b(nP)$ production might be enhanced in the
region of the $\Upsilon(5S)$, which may exist an exotic resonance
$Y_b$ analogue of the $Y(4260)$.

\begin{figure}[ht]
\begin{center}
\includegraphics[width=0.6\columnwidth]{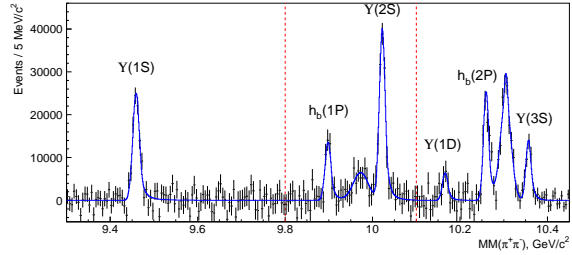}
\caption{The $MM(\pi^+\pi^-)$ spectrum with the combinatorial
background and
  $K_s$ contributions subtracted (dots with error bars) and signal
  component of the fit function (solid histogram). The vertical dashed
  lines indicate the boundaries of the fit regions. }
  \label{fig:hb-belle}
\end{center}
\end{figure}

Using the full $\Upsilon(5S)$ data sample with the integrated
luminosity of 121.1 fb$^{-1}$ collected near the peak of
$\Upsilon(5S)$ with the Belle detector, they observe the $h_b(1P)$
and $h_b(2P)$ in the missing mass spectrum of $\pi^+\pi^-$ pairs.
The $\pi^+\pi^-$ missing mass is defined as $ MM(\pi^+\pi^-) \equiv
\sqrt{(E_{c.m.}-E_{\pi^+\pi^-}^*)^2-p_{\pi^+\pi^-}^{*2}}, $ where
$E_{c.m.}$ is the center-of-mass (c.m.) energy, $E_{\pi^+\pi^-}^*$
and $p_{\pi^+\pi^-}^*$ are the $\pi^+\pi^-$ energy and momentum
measured in c.m. frame. The details of the analysis can be found
in~\cite{mizprl,bondar}.  The $MM(\pi^+\pi^-)$ spectrum with the
combinatorial background and $K_s$ contributions subtracted, and the
signal function resulting from the fit overlaid, are shown in
Fig.~\ref{fig:hb-belle}.
The significance of the $h_b(1P)$ and $h_b(2P)$ signals which
includes the systematic uncertainty is $5.5\sigma$ and $11.2\sigma$,
respectively. This is the first observation of the $h_b(1P)$ and
$h_b(2P)$ spin-singlet bottomonium states in the process $e^+e^-
\rightarrow h_b(nP) \pi^+\pi^-$ at the $\Upsilon(5S)$ energy. The
measured masses and cross-sections relative to the $e^+e^-
\rightarrow \Upsilon(2S) \pi^+\pi^-$ cross-section are
$M=9898.25\pm1.06^{+1.03}_{-1.07}$ MeV/$c^2$,
$R=0.407\pm0.079^{+0.043}_{-0.076}$ for the $h_b(1P)$ and
$M=10259.76\pm0.64^{+1.43}_{-1.03}$ MeV/$c^2$,
$R=0.78\pm0.09^{+0.22}_{-0.10}$ for the $h_b(2P)$. The mass of the
spin-singlet state is consistent with the center-of-gravity of the
corresponding $\chi_{bJ}$ states. The hyperfine splitting is $\Delta
M (1P)_{HF} = 1.62\pm1.52 $ MeV/$c^2$ ($\Delta M (2P)_{HF} =
0.48^{+1.57}_{-1.22} $ MeV/$c^2$).

\begin{figure}[ht]
\begin{center}
\includegraphics[width=0.3\columnwidth]{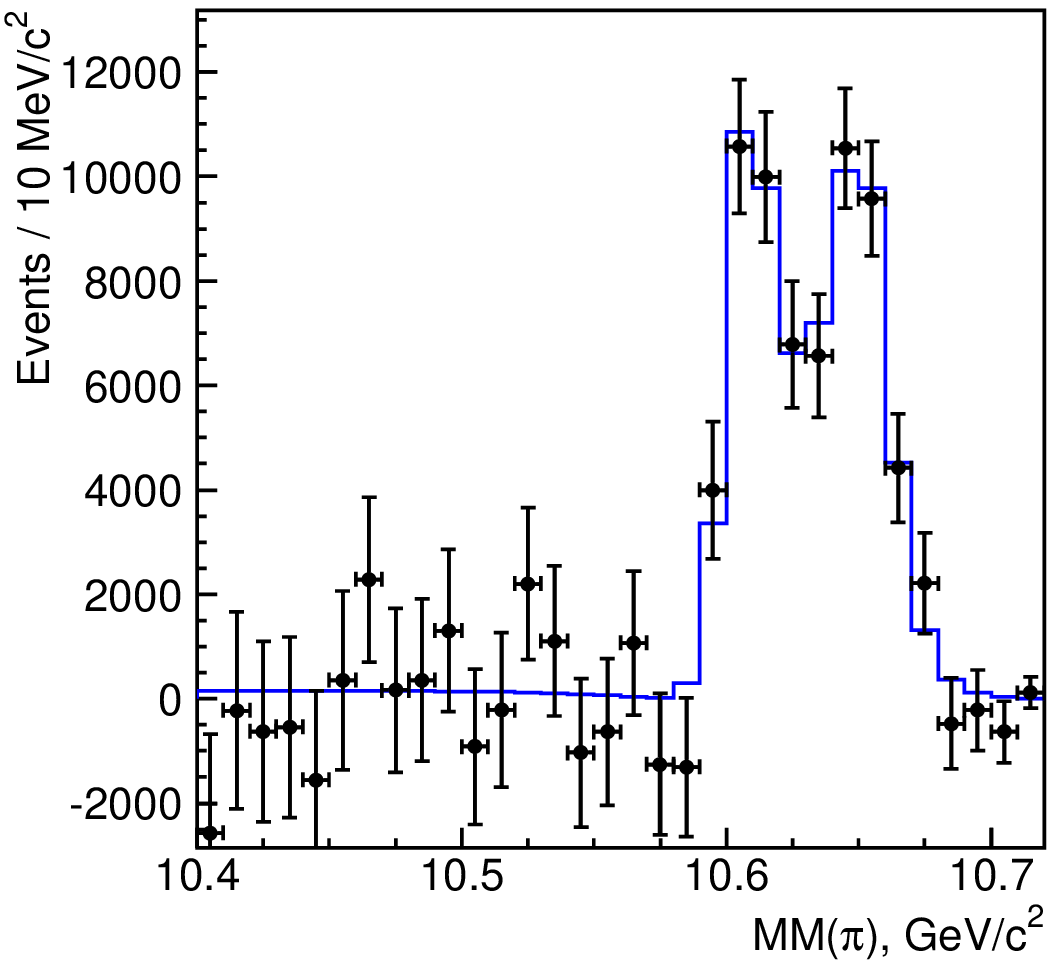}
\includegraphics[width=0.3\columnwidth]{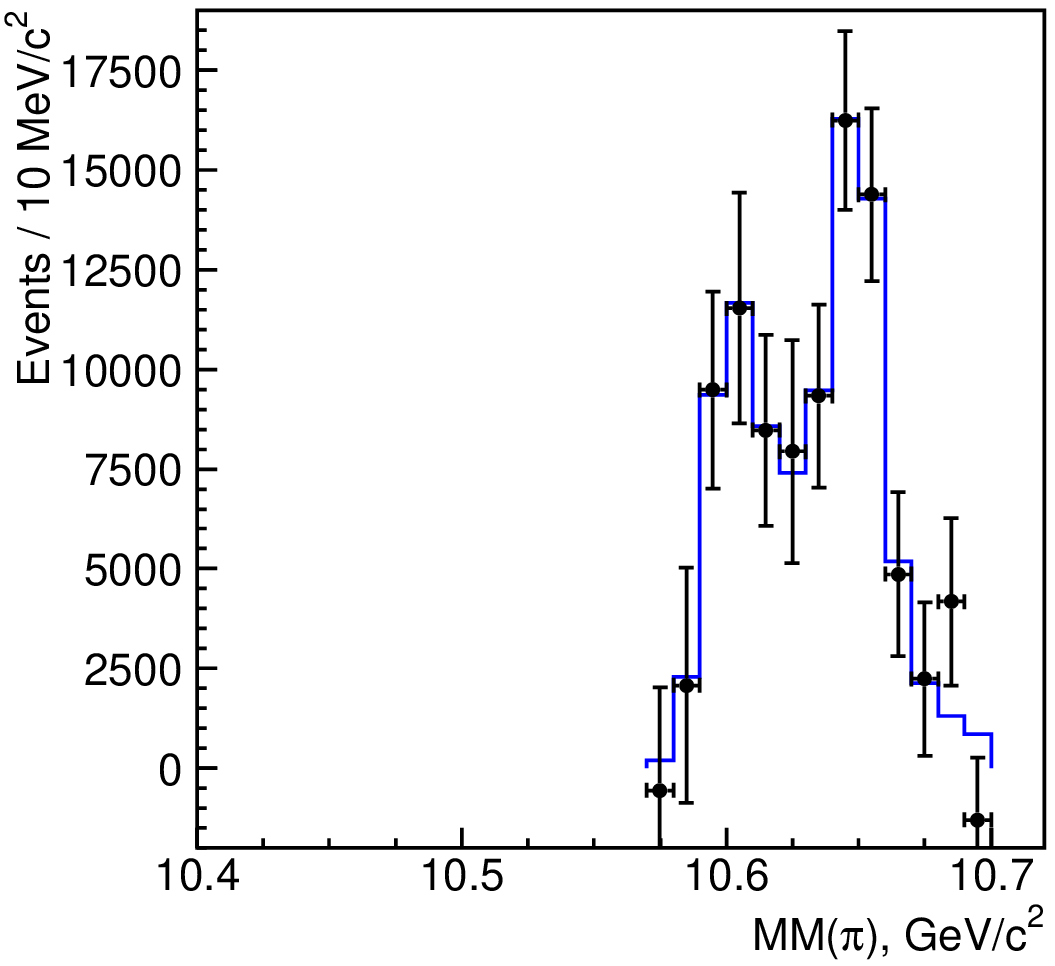}
\caption{ Left:  the yield of the $h_b(1p)$ as a
  function of $MM(\pi)$ (points with error bars) and results of the fit
  (histogram).  Right: the yield of the $h_b(2P)$ as a function of $MM(\pi)$
  (points with error bars) and results of the fit (histogram).}
  \label{fig:zb-hb}
\end{center}
\end{figure}

The large $R$ values indicate that an exotic state, $Y_b$,  may be
around $\Upsilon(5S)$ energy, and the exotic decay violates   the
suppression of heavy quark spin-flip.  Belle experiment studied the
resonant substructure of $\Upsilon(5S) \rightarrow h_b(nP)
\pi^+\pi^-$ decays~\cite{mizprl}. Since the low statistics and high
background, a Dalitz plot analysis is impossible.  Belle studied
performed the one-dimensional distribution of $M(h_b(nP)\pi)$, where
$M(h_b(nP)\pi^+)$ ($M(h_b(nP)\pi^-)$) is defined as  a missing mass
of the opposite-sign pion, $MM(\pi^-)$ ($MM(\pi^+)$).  The $h_b(nP)$
signal yields are measured as a function of the $MM(\pi^\pm)$ by
fitting the $\mmpp$ spectra in the bins of $MM(\pipm)$.
Figure~\ref{fig:zb-hb} (a) and (b) show results of the fits for the
$h_b(1P)$ and $h_b(2P)$ yields as a function of $MM(\pi)$. A clear
two-peak structure is seen without any significant non-resonant
contribution. By assuming that the spin-parity for both structures
is $J^P = 1^+$, a $\chi^2$ fit to the $MM(\pi)$ distributions is
performed. In the fit, two $P$-wave BW amplitudes and a non-resonant
contribution is used. The results of the fit are shown in
Fig.~\ref{fig:zb-hb} and are summarized in Table~\ref{tab:zb-hb}.
The non-resonant amplitude is found to be consistent with zero. In
the fit,  the hypothesis of two resonances is favored over the
hypothesis of a single resonance (no resonances) at the
$7.4\,\sigma$ ($17.9\,\sigma$) level.  Therefore, the two charged
structures are named as $Z_b(10610)$ and $Z_b(10650)$, respectively.
The parameters of the  $Z_b(10610)$ and $Z_b(10650)$ obtained in the
fit of $h_b(1P)$ and $h_b(2P)$ are consistent with each other.
\begin{table}[!t]
  \caption{Comparison of results on $Z_b(10610)$ and $Z_b(10650)$ parameters
           obtained from
            $\Upsilon(5S) \rightarrow  h_b(nP) \pi^+\pi^-$
           ($n=1,2$) analyses. Quoted values are in MeV/$c^2$ for masses, in
           MeV for widths and in degrees for the relative phase. Relative
           amplitude is defined as $a_{Z_b(10650)}/a_{Z_b{10610}}$.}
  \medskip
  \label{tab:zb-hb}
\centering { \scriptsize
  \begin{tabular}{lcc} \hline \hline
 Final state &
               $\hb\pp$                   &
               $\hbp\pp$
\\ \hline
           $M(Z_b(10610))$ &
           $\mzahb$          &
           $\mzahbp$
 \\
           $\Gamma(Z_b(10610))$ &

           $\gzahb$   &
           $\gzahbp$
 \\
           $M(Z_b(10650))$ &
           $\mzbhb$      &
           $\mzbhbp$
 \\
           $\Gamma(Z_b(10650))$ &
           $\gzbhb$           &
           $\gzbhbp$
 \\
           Rel. amplitude                 &
           $\ahb$    &
           $\ahbp$
 \\
           Rel. phase, &
           $\phihb$         &
           $\phihbp$
\\
\hline \hline
\end{tabular}}
\end{table}

\begin{table}[!t]
  \caption{Comparison of results on $Z_b(10610)$ and $Z_b(10650)$ parameters
           obtained from $\Upsilon(5S) \rightarrow \Upsilon(nS)\pi^+\pi^-$ ($n=1,2,3$) analyses. Quoted values are in MeV/$c^2$ for masses, in
           MeV for widths and in degrees for the relative phase. Relative
           amplitude is defined as $a_{Z_b(10650)}/a_{Z_b{10610}}$.}
  \medskip
  \label{tab:zb-yb}
\centering { \scriptsize
  \begin{tabular}{lccc} \hline \hline
 Final state & $\Uo\pp$                   &
               $\Ut\pp$                   &
               $\Uth\pp$
\\ \hline
           $M(Z_b(10610))$ &
           $10609\pm3\pm2$                &
           $10616\pm2^{+3}_{-4}$          &
           $10608\pm2^{+5}_{-2}$
 \\
           $\Gamma(Z_b(10610))$ &
           $22.9\pm7.3\pm2$               &
           $21.1\pm4^{+2}_{-3}$           &
           $12.2\pm1.7\pm4$
 \\
           $M(Z_b(10650))$ &
           $10660\pm6\pm2$                &
           $10653\pm2\pm2$                &
           $10652\pm2\pm2$
 \\
           $\Gamma(Z_b(10650))$ &
           $12\pm10\pm3$~                 &
           $16.4\pm3.6^{+4}_{-6}$         &
           $10.9\pm2.6^{+4}_{-2}$
 \\
           Rel. amplitude                 &
           $0.59\pm0.19^{+0.09}_{-0.03}$  &
           $0.91\pm0.11^{+0.04}_{-0.03}$  &
           $0.73\pm0.10^{+0.15}_{-0.05}$
 \\
           Rel. phase, &
           $53\pm61^{+5}_{-50}$           &
           $-20\pm18^{+14}_{-9}$          &
           $6\pm24^{+23}_{-59}$
\\
\hline \hline
\end{tabular}}
\end{table}

\begin{figure}[ht]
  \centering
\hspace*{-1mm}
  \includegraphics[width=0.3\textwidth]{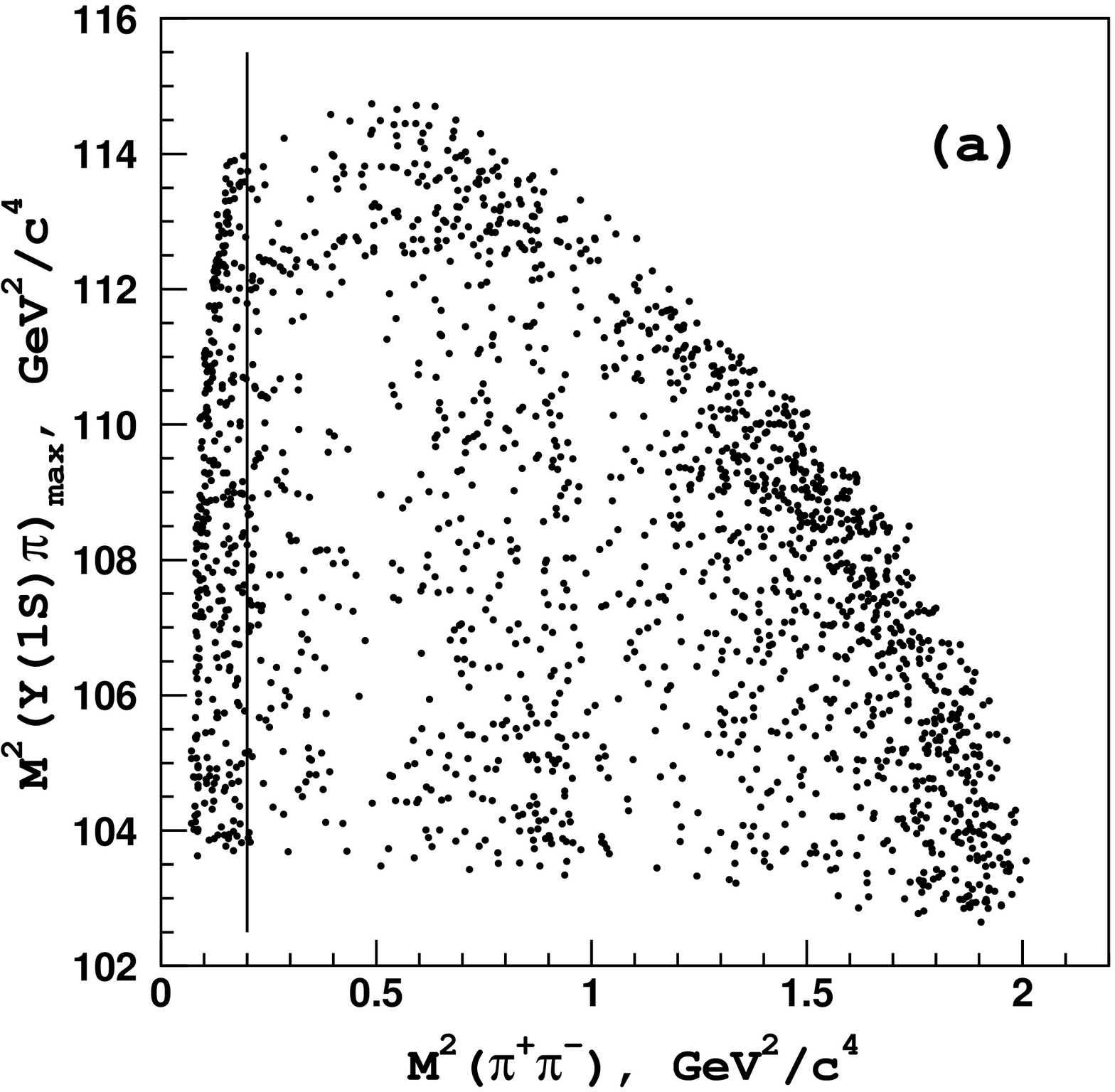} \hfill
  \includegraphics[width=0.3\textwidth]{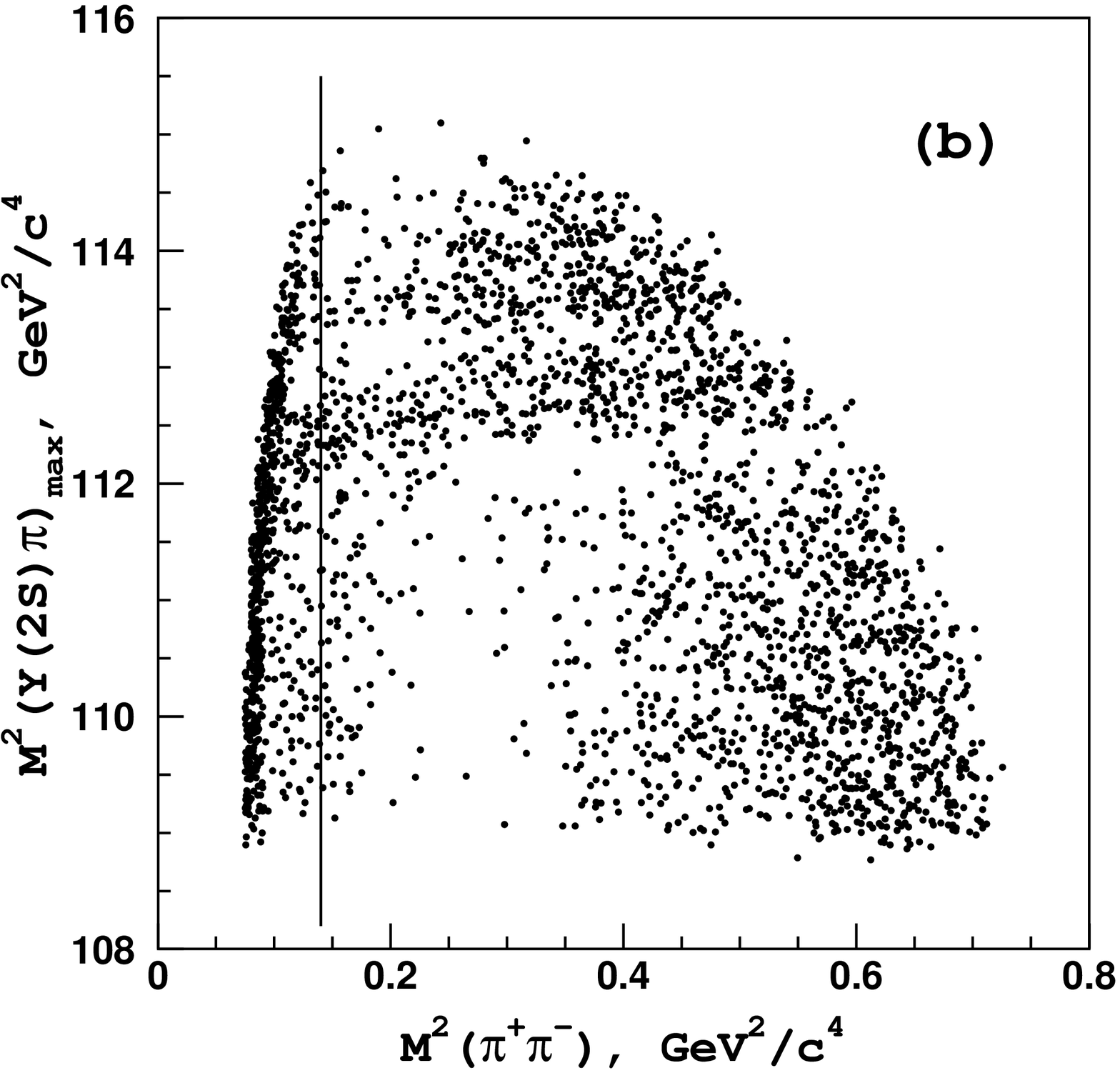} \hfill
  \includegraphics[width=0.3\textwidth]{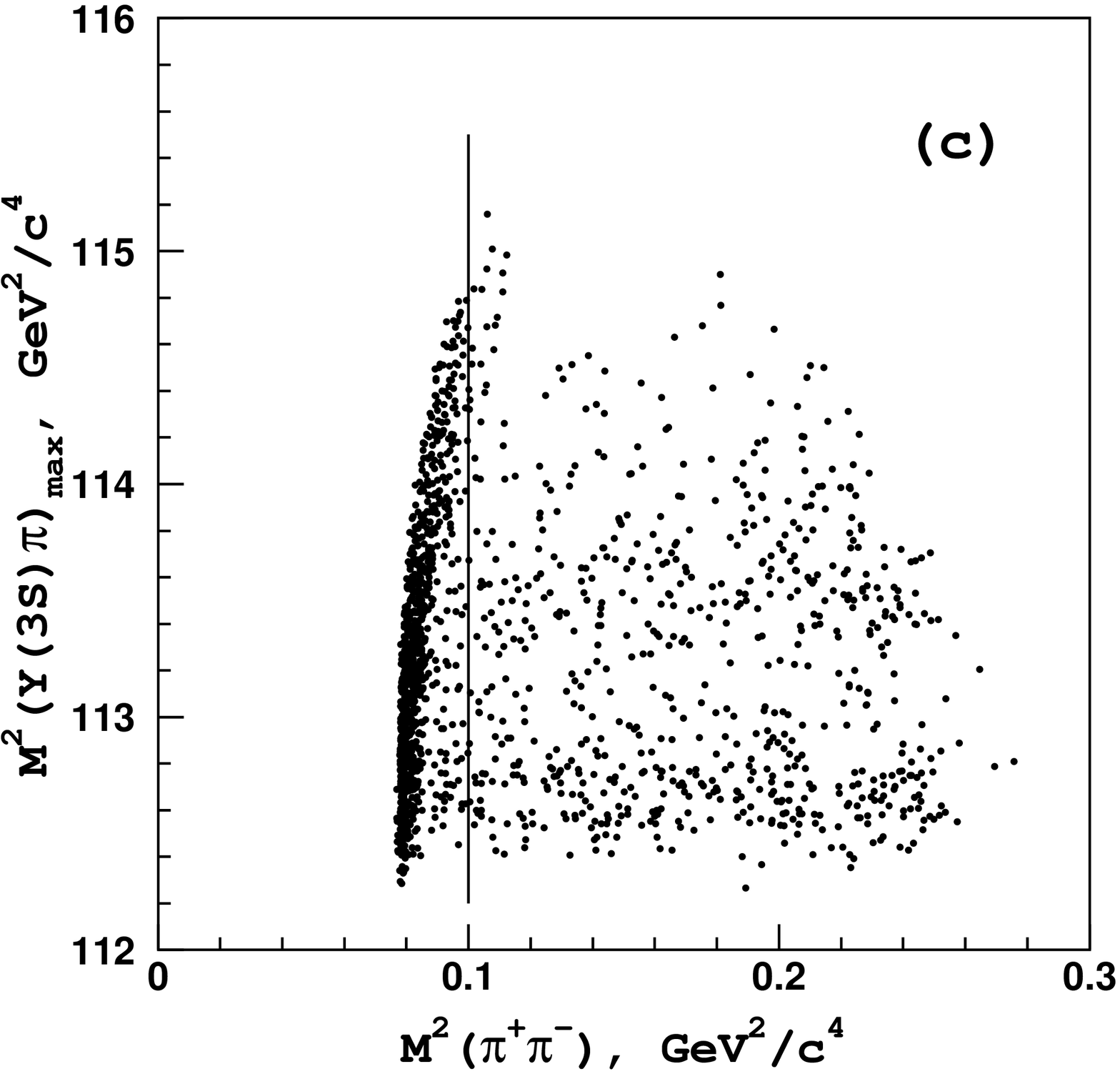}
  \caption{Dalitz plots for $\Un\pp$ events in the (a) $\Uo$; (b)
    $\Ut$; \mbox{(c) $\Uth$} signal regions. Dalitz plot regions to
    the right of the vertical lines are included in the amplitude
    analysis.}
\label{fig:ynspp-s-dp}
\end{figure}

\begin{figure}[ht]
  \centering
  \includegraphics[width=0.3\textwidth]{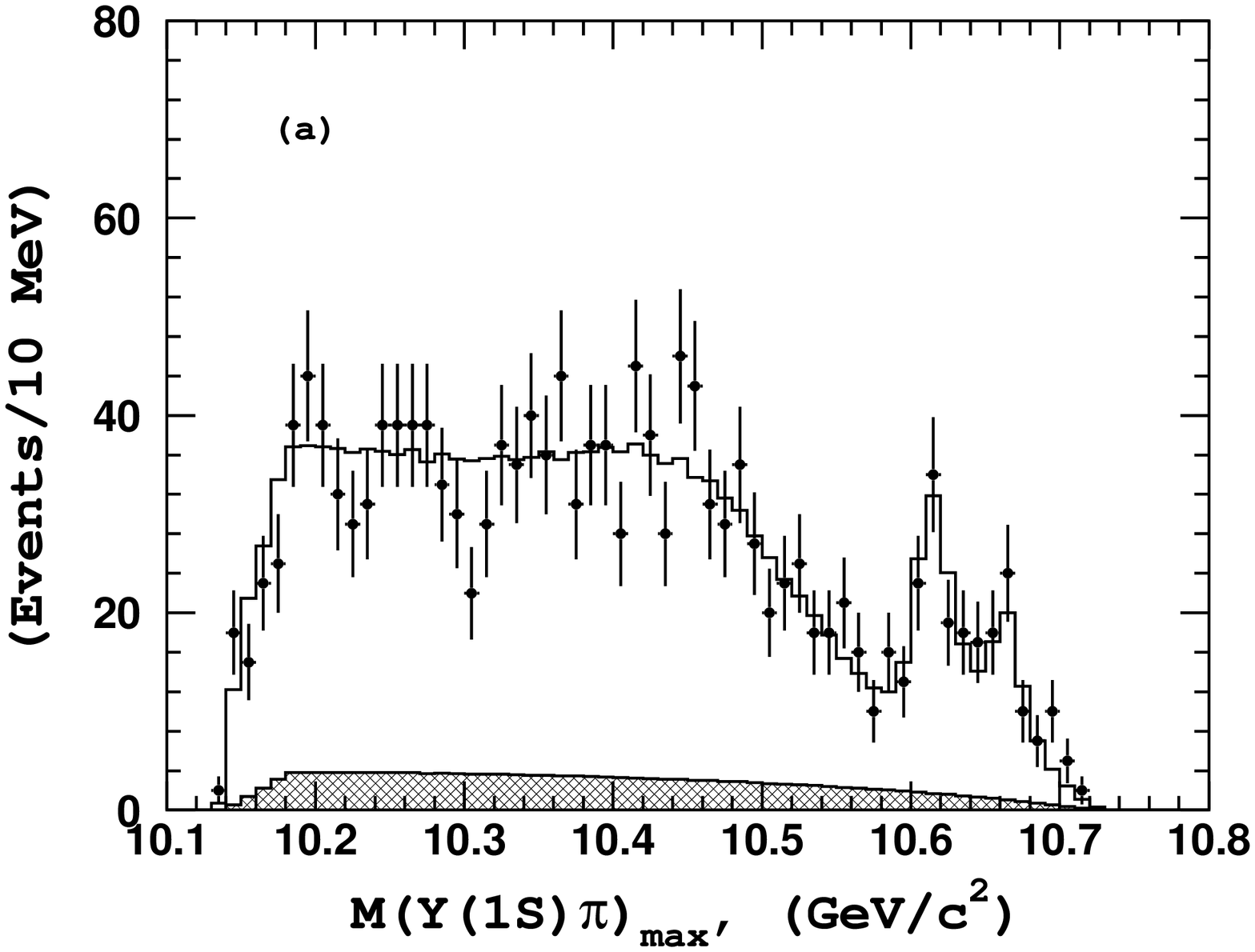} \hfill
  \includegraphics[width=0.3\textwidth]{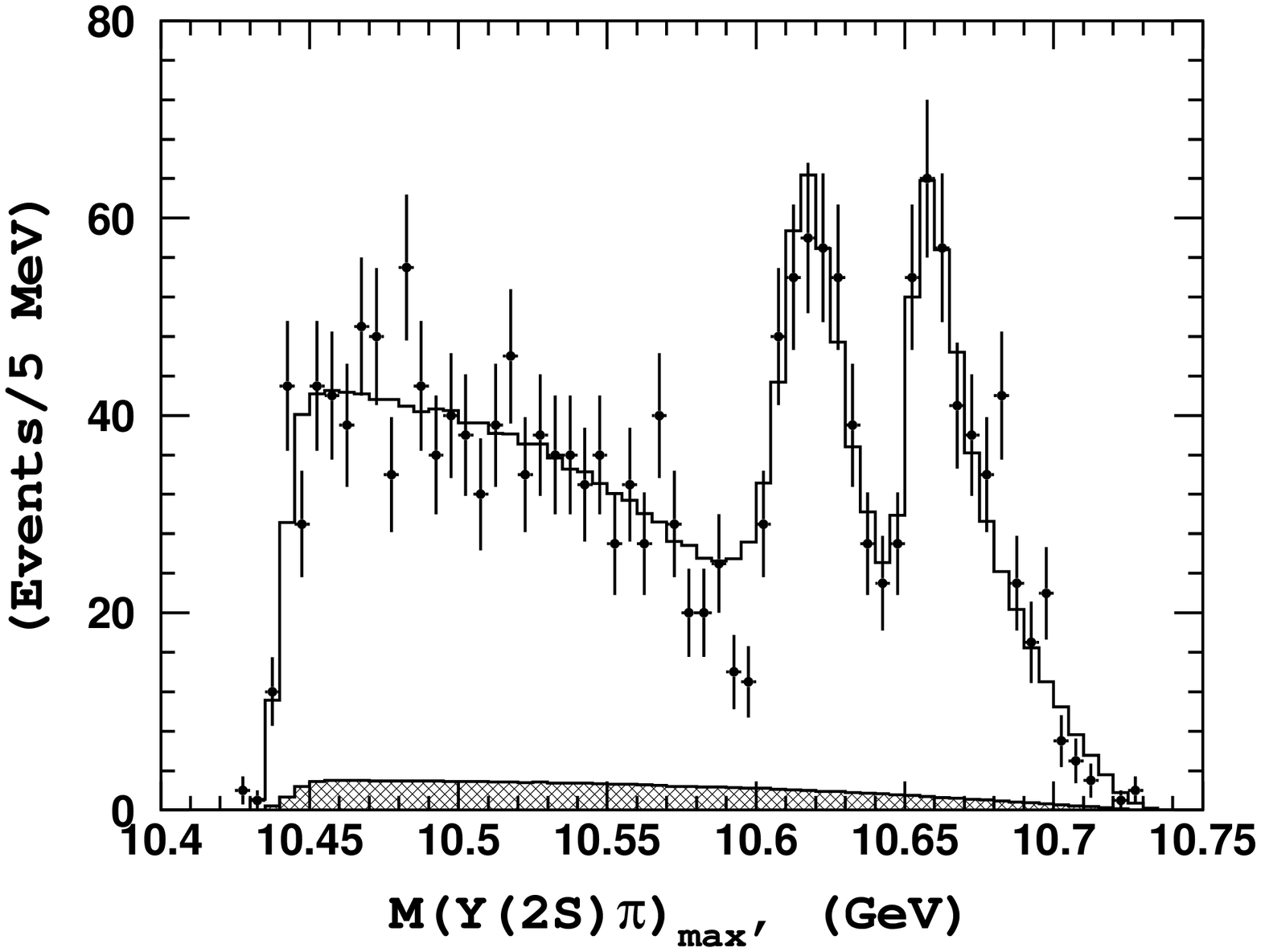} \hfill
  \includegraphics[width=0.3\textwidth]{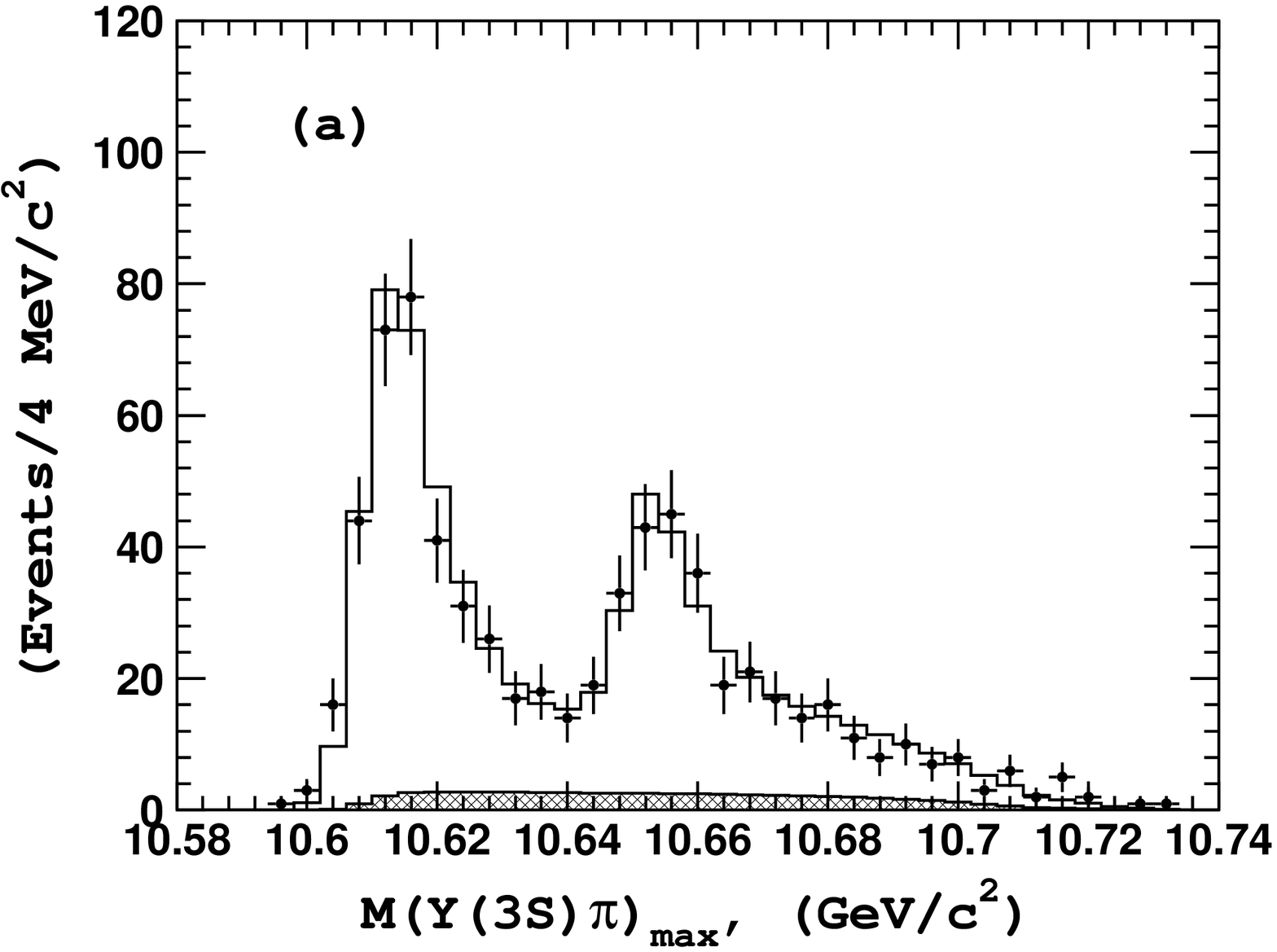} \\
  \caption{Comparison of fit results (open histogram) with
    experimental data (points with error bars) for events in the $\Uo$, $\Ut$ and $\Uth$  signal
    regions.  The hatched histogram shows the background component.}
\label{fig:y3spp-f-hh}
\end{figure}

 For the analysis of the $\Upsilon(5S) \rightarrow \Upsilon(nS)
 \pi^+\pi^-$, in addition of the the missing
mass $MM(\pi^+\pi^-)$ associated with the $\pp$ system, the events
are further identified by the $\mu^+\mu^-$
 pair with an invariant mass in the range of
$8.0~\gevm<M(\mu^+\mu^-)<11.0~\gevm$. Figure~\ref{fig:ynspp-s-dp}
shows Dalitz plots of the events in the signal regions for the three
decay channels under study. In all cases, two horizontal bands are
evident in the $\Un\pi$ system near $10.61\,\gevm$
($\sim112.6$~GeV$^2/c^4$) and $10.65\,\gevm$
($\sim113.3$~GeV$^2/c^4$). The amplitude analyses of the three-body
$\Uf\to\Un\pp$ decays that are reported here are performed by means
of unbinned maximum likelihood fits to two-dimensional Dalitz
distributions.
The details of the description of the Dalitz plot analysis can be
found in~\cite{bondar,belle-yb}.  Results of the fits to
$\Uf\to\Un\pp$ signal events are shown in Fig.~\ref{fig:y3spp-f-hh},
where one-dimensional projections of the data and fits are compared.
To combine $Z^+_b$ and $Z^-_b$ events we plot $\Un\pi$ mass
distributions in terms of $M(\Un\pi)_{\min}$ and $M(\Un\pi)_{\max}$;
fits are performed in terms of $M(\Un\pi^+)$ and $M(\Un\pi^-)$.
Results of the fits are summarized in Table~\ref{tab:zb-yb}. The
combined statistical significance of the two peaks exceeds
10$\sigma$ for all tested models and for all $\Un\pp$ decay modes.
\begin{figure}[ht]
\centering
\includegraphics[width=0.60\textwidth]{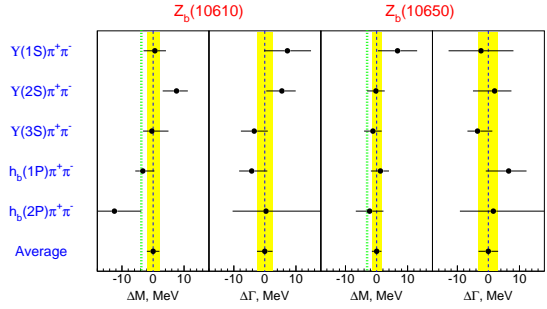}
\caption{Comparison of $Z_b(10610)$ and $Z_b(10650)$ parameters
  obtained from different decay channels. The vertical dotted lines
  indicate $B^*{\overline B}$ and $B^*{\overline B^*}$ thresholds.}
\label{fig:summary}
\end{figure}

The Belle experiment observed two charged bottomonium-like
resonances, the $\zbo$ and $\zbt$, with signals in five different
decay channels, $\Un\pipm$ ($n=1,2,3$) and $\hbn\pipm$ ($m=1,2$).
All channels yield consistent results as can be seen in
Fig.~\ref{fig:summary}.  A simple weighted averages over all five
channels give $M[Z_b(10610)]=10608.4\pm2.0\,\mevm$,
$\Gamma[Z_b(10610)]=15.6\pm2.5\,\mev$ and
$M[Z_b(10650)]=10653.2\pm1.5\,\mevm$,
$\Gamma[Z_b(10650)]=14.4\pm3.2\,\mev$, where statistical and
systematic errors are added in quadrature. The measured masses of
these states exceed by only a few MeV/$c^2$ the thresholds for the
open beauty channels $B^*{\overline B}$ ($10604.6$~MeV) and $B^*
{\overline B^*}$ ($10650.2$~MeV).  This ``coincidence'' can be
explained by a molecular-like type of new states, {\it i.e.}, their
structure is determined by the strong interaction dynamics of the
$B^* {\overline B}$ and $B^*{\overline
  B^*}$ meson pairs~\cite{bondar-1}.
The $\Uf\to\hbn\pp$ decays seem to be saturated by the $\zbo$ and
$\zbt$ intermediate states; this decay mechanism may be responsible
for the high rate of the $\Uf\to\hbn\pp$ process measured recently
by the Belle Collaboration~\cite{belle-y-decay}.

\section{Summary}

In this paper, I review the most recent experimental results on the
light hadron, charmonium and bottomonium states. Especially, I gave
a review of the exotic candidates observed so far. Although a lot of
great progresses have been made in the last few years, the situation
is far from being completely clarified. In most case, new data and
high statistics are needed, and they are expected to come from the
LHC and the future flavor factories. At the same time, the analysis
effort of the B-Factory experiments is still on going and new
results are expected to come in the near future. Finally, we need
more theoretical efforts to make further investigations, for
example, in the case of $\Upsilon(5S) \rightarrow
\Upsilon(nS)\pi^+\pi^-$ and $h_b(nS) \pi^+\pi^-$ interpretation and
their connections with the $e^+e^- \rightarrow \text{hadrons}$ cross
sections. In conclusion, hadron spectroscopy is still an intriguing
field, new or more precise measurements will continue to provide, in
the near future, important information to better understand QCD and
its effective treatments, with a broad impact on many other fields.

\acknowledgments

The author would like to thank experimental colleagues from BESIII,
BABAR, Belle, CLEO, CDF, LHCb, CMS, KLEO and KEDR Collaborations.
Especially, thanks to Hesheng Chen, Yifang Wang, Steve Olsen,
X.Y.Shen, C.Z.Yuan, Torsten Schroer, Karim Trabelsi, A. E. Bondar,
S.-K.Choi, Roy A. Briere, K. Seth, M. Kornicer, Diego Tonelli,
Giovanni Punzi, Robert Harr, Y.N. Gao, Hal Evans, Simon I. Eidelman,
Evgeny Baldin, Simona Giovannella and Paolo Gauzzi. I am grateful
for support from my institute and from National Natural Science
Foundation of China under contract No. 11125525.


\end{document}